\documentclass[a4paper,prd,nofootinbib,preprint,showpacs,showkeys]{revtex4}

\usepackage[english]{babel}
\usepackage{amsmath,amssymb}
\usepackage[dvips]{graphicx}

\renewcommand{\onlinecite}{\cite}

\allowdisplaybreaks

\DeclareMathAlphabet{\mathsc}{OT1}{cmr}{m}{sc}

\newcommand{\CL} {C.L.}
\newcommand{\dof}{d.o.f.}
\newcommand{\eVq}{\text{eV}^2}

\newcommand{\Sol}  {\mathsc{sol}}
\newcommand{\Atm}  {\mathsc{atm}}
\newcommand{\Sbl}  {\mathsc{sbl}}
\newcommand{\Nev}  {\mathsc{nev}}
\newcommand{\Lsnd} {\mathsc{lsnd}}
\newcommand{\Chooz}{\mathsc{chooz}}

\newcommand{\pnu}[1] {\overset{\smash{\scriptscriptstyle (-)}}{\nu}_{\hskip-3pt #1}}
\newcommand{\dms}{\Delta m^2_\Sol}
\newcommand{\dma}{\Delta m^2_\Atm}
\newcommand{\dml}{\Delta m^2_\Lsnd}
\newcommand{\thl}{\theta_\Lsnd}
\newcommand{\sql}{\sin^2 2\thl}

\newcommand{\myodot}{{\mathchoice
    {\raisebox{\depth}{$\displaystyle\odot$}}
    {\raisebox{\depth}{$\textstyle\odot$}}
    {\raisebox{\depth}{$\scriptscriptstyle\odot$}}
    {\raisebox{\depth}{$\scriptscriptstyle\odot$}}
    }}

\begin{document}

\preprint{hep-ph/0112103}
\preprint{IFIC/01--67}

\title{\vspace*{1 cm}Status of four-neutrino mass schemes: a global
  and unified approach to current neutrino oscillation data}

\author{M.~Maltoni} \email{maltoni@ific.uv.es}
\author{T.~Schwetz} \email{schwetz@ific.uv.es}
\author{J.~W.~F.~Valle} \email{valle@ific.uv.es}
\affiliation{Instituto de F\'{\i}sica Corpuscular --
  C.S.I.C./Universitat de Val{\`e}ncia \\
  Edificio Institutos de Paterna, Apt 22085,
  E--46071 Valencia, Spain\vspace{2 cm}}

\begin{abstract}
    We present a unified global analysis of neutrino oscillation data
    within the framework of the four-neutrino mass schemes (3+1) and
    (2+2). We include all data from solar and atmospheric neutrino
    experiments, as well as information from short-baseline
    experiments including LSND.
    If we combine only solar and atmospheric neutrino data, (3+1)
    schemes are clearly preferred, whereas short-baseline data in
    combination with atmospheric data prefers (2+2) models.
    When combining all data in a global analysis the (3+1) mass scheme
    gives a slightly better fit than the (2+2) case, though all
    four-neutrino schemes are presently acceptable. The LSND result
    disfavors the three-active neutrino scenario with only $\dms$ and
    $\dma$ at 99.9\% \CL\ with respect to the four-neutrino best fit
    model.
    We perform a detailed analysis of the goodness of fit to identify
    which sub-set of the data is in disagreement with the best fit
    solution in a given mass scheme.
\end{abstract}

\keywords{neutrino oscillations, sterile neutrino, four-neutrino models}
\pacs{14.60.P, 14.60.S, 96.40.T, 26.65, 96.60.J, 24.60}

\maketitle

\section{Introduction}

From the long-standing solar~\cite{sksol,chlorine,sage,gallex_gno,sno}
and atmospheric~\cite{atm-exp,skatm,macroOld,macroNew} neutrino
anomalies we now have compelling evidence that an extension of the
Standard Model of particle physics is necessary in the lepton sector.
The most natural explanation of these experiments is provided by
neutrino oscillations induced by neutrino masses and mixing with
neutrino mass-squared differences of the order of $\dms\lesssim
10^{-4}~\eVq$ and $\dma\sim 3\times 10^{-3}~\eVq$. Explaining also the
evidence of $\pnu{\mu} \to \pnu{e}$ oscillations with a mass-squared
difference $\dml \sim 1~\eVq$ reported by the LSND
experiment~\cite{LSND,LSND2001} requires an even more radical
modification of the Standard Model. Currently this experiment is left
out in most analyses of neutrino data. At the moment the LSND result
is not confirmed nor ruled out by any other experiment, and therefore
it is reasonable to see more quantitatively its impact on the physics
of the lepton sector.

If all the three anomalies are explained by neutrino oscillations, and
the possibility of CPT violation is neglected~\cite{CPT}, we need at
least four neutrinos to obtain the three required mass-squared
differences. In view of the LEP results the fourth neutrino must not
couple to the Z-boson. Such a {\it sterile neutrino} with a mass in
the electron-volt range has been postulated originally to provide some
hot dark matter suggested by early COBE
results~\cite{ptv92,pv93,cm93}, and after the LSND result many
four-neutrino models have been
proposed~\cite{4nuModels,4nuextra,Ioannisian:2001mu,Hirsch:2000xe}.  A
quite complete list on four-neutrino references can be found
at~\cite{giuntiwebp}.

A very important issue in the context of four-neutrino scenarios is
the question of the four-neutrino mass spectrum.  Two very different
classes of four-neutrino mass spectra can be identified.  The first
class contains four types and consists of spectra where three neutrino
masses are clustered together, whereas the fourth mass is separated
from the cluster by the mass gap needed to reproduce the LSND result.
The second class has two types where one pair of nearly degenerate
masses is separated by the LSND gap from the two lightest neutrinos.
These two classes are referred to as (3+1) and (2+2) neutrino mass
spectra, respectively~\cite{barger00}. All possible four-neutrino mass
spectra are shown in Fig.~\ref{fig:4spectra}.

One important theoretical issue in these models is how to account for
the lightness of the sterile neutrino which, ordinarily, should have
mass well above the weak scale. The simplest possibility is to appeal
to an underlying protecting symmetry, getting, moreover, the LSND mass
at one-loop order only~\cite{ptv92,pv93}.
Alternatively, the lightness of the sterile neutrino may follow from
volume suppression in models based on extra
dimensions~\cite{4nuextra,Ioannisian:2001mu}.
As for the maximal atmospheric mixing angle, it follows naturally in
the models of Refs.~\onlinecite{ptv92,pv93,Ioannisian:2001mu} since to
first approximation the heaviest neutrinos form a quasi-Dirac pair
whose components mix maximally.
Finally the splittings which generate solar and atmospheric
oscillations arise due to breaking of the original symmetry (for
example due to additional loop suppression)~\cite{ptv92,pv93} or due
to R-parity breaking~\cite{Hirsch:2000xe}. These models lead to a
(2+2) scheme.

One important feature of (3+1) mass spectra is that they include the
three-active neutrino scenario as limiting case. In this case solar
and atmospheric neutrino oscillations are explained by active neutrino
oscillations, with mass-squared differences $\dms$ and $\dma$, and the
fourth neutrino state gets completely decoupled.  We will refer to
this scenario as (3+0). The (3+1) scheme can be considered as a
perturbation of the (3+0) case: a small mixture of $\nu_e$ and
$\nu_\mu$ with the separated mass state can account for the
oscillations observed by LSND. In contrast, the (2+2) spectrum is
intrinsically different from the three-active neutrino case. A very
important prediction of this mass spectrum is that there has to be a
significant contribution of the sterile neutrino either in solar or in
atmospheric neutrino oscillations or in both.  More precisely, in the
(2+2) case the fractions of sterile neutrino participating in solar
and in atmospheric oscillations have to add up to one~\cite{peres}.

Based on semi-quantitative arguments it has been realized for some
time~\cite{OY,barger98,BGGS,BGG} that it is difficult to explain the
LSND result in the framework of (3+1) schemes because of strong bounds
from negative neutrino oscillation searches in short-baseline (SBL)
experiments, and therefore the (2+2) scheme was considered as the
preferred one. Recent experimental developments lead to a renaissance
of the (3+1) mass schemes~\cite{barger00,peres,carlo}.  First, a new
LSND analysis (see last reference of~\cite{LSND}) resulted in a shift
of the region allowed by LSND to slightly smaller values of $\dml$,
which makes the (3+1) schemes somewhat less disfavored. However, in
Refs.~\onlinecite{GS,Maltoni:2001mt} it was shown within a well
defined statistical analysis that a bound implied by SBL experiments
is in disagreement even with the new LSND allowed region at the 95\%
\CL\ in (3+1) schemes.  Second, the high statistics data from
Super-Kamiokande started to exclude two-neutrino oscillations into a
sterile neutrino for both solar as well as atmospheric
neutrinos~\cite{sksterile}, which constitutes a problem for (2+2) mass
schemes.  Concerning the solar data, the trend to disfavor
oscillations into a sterile neutrino recently became supported by the
beautiful result of the SNO experiment~\cite{sno,bahcall}. However, a
unified analysis of solar and atmospheric neutrino data performed in
Refs.~\onlinecite{Concha:2001uy,Concha:2001zi} showed that the
goodness of fit of the (2+2) mass scheme is still acceptable.

In this work we perform for the first time a global analysis of all
the relevant neutrino oscillation data in a four-neutrino framework.
We will use the fit of the global solar neutrino data presented in
Ref.~\onlinecite{Concha:2001zi}, which includes
Super-Kamiokande~\cite{sksol}, Homestake~\cite{chlorine},
SAGE~\cite{sage}, GALLEX and GNO~\cite{gallex_gno} and SNO~\cite{sno}.
Further we include data from the atmospheric neutrino experiments
Super-Kamiokande~\cite{skatm} and MACRO~\cite{macroOld},
data from the SBL $\pnu{\mu} \to \pnu{e}$ appearance
experiments LSND~\cite{LSND2001}, KARMEN~\cite{KARMEN} and
NOMAD~\cite{nomad}, the reactor $\bar\nu_e$ disappearance experiments
Bugey~\cite{bugey} and CHOOZ~\cite{CHOOZ} and the $\pnu{\mu}$
disappearance experiment CDHS~\cite{CDHS}. We will perform a fit to
these data for (3+1) and (2+2) mass spectra in a unified formalism,
which allows to compare directly the quality of the fit for these
rather different mass schemes.

The plan of the paper is as follows. In Sec.~\ref{sec:parameters} we
define our notation. In Secs.~\ref{sec:SBL}, \ref{sec:solar} and
\ref{sec:atmospheric} we consider the mixing parameters relevant in
the different classes of experiments (SBL, solar and atmospheric,
respectively), discuss constraints on these parameters and describe
the experimental data used in the analysis. In
Sec.~\ref{sec:paramSummary} we give a thorough discussion of the
parametrization of four-neutrino mass schemes. The parameterization we
introduce is based on physically relevant quantities and is convenient
for the combined analysis. In Sec.~\ref{sec:sol+atm} we compare the
(3+1) and (2+2) mass schemes when data from solar and atmospheric
neutrino experiments are combined, whereas in Sec.~\ref{sec:atm+SBL}
we consider the combination of atmospheric and SBL experiments. In
Sec.~\ref{sec:global} we present the main result of this work: a fit
of the global set of neutrino oscillation data (solar, atmospheric and
SBL) in the framework of four-neutrino mass spectra.  We also comment
on a Standard Model fit, {\it i.e.}\ a fit for the (3+0) case.  In
Secs.~\ref{sec:sol+atm}--\ref{sec:global} we focus mainly on the
relative comparison of (3+1) and (2+2) mass schemes; we will make some
comments on the absolute quality of the fit in Sec.~\ref{sec:gof}.
Finally, we present our conclusions in Sec.~\ref{sec:conclusions}.

For readers interested mainly in the results of our work we suggest to
skip Secs.~\ref{sec:SBL}--\ref{sec:paramSummary}. After having a look
at Fig.~\ref{fig:parameters}, where the parameter structure of the
four-neutrino fit is illustrated, we recommend to proceed directly to
Sec.~\ref{sec:sol+atm}.

\section{Four-neutrino oscillation parameters}
\label{sec:parameters}

\begin{figure}[t]
   \includegraphics[width=0.75\linewidth]{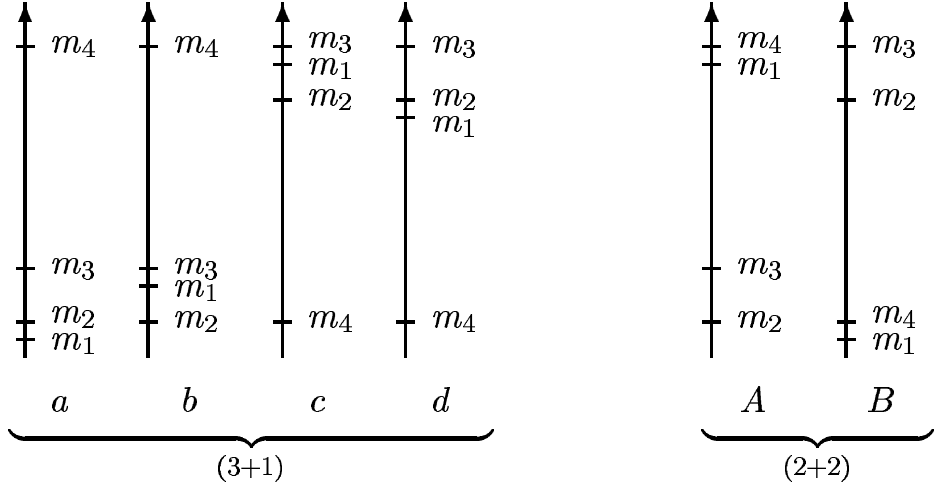}
      \caption{\label{fig:4spectra}%
      The six types of four-neutrino mass spectra. The different
      distances between the masses on the vertical axes symbolize the
      different scales of mass-squared differences required to explain
      solar, atmospheric and LSND data with neutrino oscillations.}
\end{figure}

To obtain a four-neutrino scenario from a gauge model of the weak
interaction one needs to extend the lepton sector by a number $m$ of
$SU(2)\otimes U(1)$ singlet leptons~\cite{Schechter:1980gr}.
In such scheme the charged current leptonic weak interaction is
specified as by a rectangular $3\times (3+m)$ lepton mixing matrix $K
= \Omega U$ which comes from diagonalizing separately the $3\times 3$
charged lepton mass matrix (via $\Omega$) as well as the, in general
$(3+m)\times (3+m)$ Majorana, neutrino mass matrix (via $U$). Moreover
the weak neutral current couplings of mass-eigenstate neutrinos is
characterized by a non-trivial $(3+m)\times (3+m)$ coupling matrix $P
= K^\dagger K$~\cite{Schechter:1980gr} whose effects will not be
relevant for us as neutrinos are both produced and detected through
charged current interactions.
If these extra singlets are all super-heavy (one example is the
standard seesaw scheme, where $m=3$), they decouple, leaving to a
nearly unitary $3\times 3$ lepton mixing matrix $K$ while the
projective $(3+m)\times (3+m)$ matrix $P$ becomes approximately the
$3\times 3$ unit matrix (approximate GIM mechanism).

From here on-wards we assume that, due to some symmetry or another
reason~\cite{ptv92,pv93,cm93,4nuextra,Ioannisian:2001mu} one of the
$SU(2)\otimes U(1)$ singlets remains light enough so that it can take
part in the oscillation phenomenology and thereby account for the LSND
data.  The minimum possibility is to have just one such light singlet,
$m=1$, called sterile neutrino.

In general the physics of four-neutrino oscillations involves 3
mass-squared differences and the elements of the mixing matrix $K$.
The latter have been characterized in a model-independent way in
\cite{Schechter:1980gr} where an explicit parametrization was given
which is, up to factor ordering, the standard one. In full generality
$K$ contains 6 mixing angles and 3 physical phases which could lead to
CP violation in the oscillation phenomena~\cite{Schechter:1981gk}.
For convenience this $3\times 4$ matrix $K$ connecting the 4 neutrino
mass fields $\nu_i$ and the 3 flavor fields $\nu_\alpha$ can be completed
with an extra line (relating the {\it sterile} neutrino $\nu_s$ to the
mass eigenstates) so to obtain a $4\times 4$ unitary matrix. In a
basis where the charged lepton mass matrix is diagonal this leads to
the matrix $U$ diagonalizing the neutrino mass matrix:
\begin{equation}
    \nu_\alpha = \sum_{i=1}^4 U_{\alpha i} \nu_i \qquad
    (\alpha = e,\mu,\tau,s) \,.
\end{equation}

Because of the strong hierarchy of the mass-squared differences
required by the experimental data the CP-violating effects are
expected to be small in the experiments we consider. However, CP
violation can be important in four-neutrino schemes for future
long-baseline experiments, such as neutrino factories~\cite{CP}. Thus,
neglecting the complex phases we are left altogether with nine
parameters relevant for the description of CP conserving neutrino
oscillations in a four-neutrino scheme: 6 mixing angles contained in
$U$ and 3 mass-squared differences.
In the following sections we will present a choice for these
parameters, which is convenient for the combined analysis of the
different experiments and which is motivated by their physical
interpretation.

We label the neutrino masses as indicated in Fig.~\ref{fig:4spectra}
and define for all schemes\footnote{Note that our labeling is
  different from the one in previous
  publications~\cite{BGGS,BGG,Maltoni:2001mt,Concha:2001uy}. However,
  this way of labeling neutrino masses is particularly convenient as it
  enables a combined treatment of all the schemes in the same footing.}
\begin{equation}
    \dml = m_4^2 - m_2^2
    \quad\text{and}\quad
    \dma = m_3^2 - m_2^2 > 0 \,.
\end{equation}
All experiments we consider are insensitive to the sign of $\dml$.
This implies that the (3+1)$_a$ scheme is equivalent to (3+1)$_d$,
while (3+1)$_b$ is equivalent to (3+1)$_c$ and (2+2)$_A$ is equivalent
to (2+2)$_B$.\footnote{These degeneracies can be lifted by considering
  the effects in tritium $\beta$-decay
  experiments~\cite{Maltoni:2001mt,farzan} or neutrino-less double
  $\beta$-decay experiments~\cite{nu0bdec}.} Hence, without loss of
generality we can restrict ourselves to the discussion of the schemes
(3+1)$_a$, (3+1)$_b$ and (2+2)$_A$, and we always have $\dml>0$.  The
structure of the neutrino mass eigenstates $\nu_2$, $\nu_3$ and
$\nu_4$ is common for all these schemes. Only the ``solar mass state''
$\nu_1$ is inserted in different places. Let us define the index
$\myodot$ for the different schemes as
\begin{equation}\label{eq:odot}
    \text{(3+1)}_a:\: \myodot \equiv 2 \,,\qquad
    \text{(3+1)}_b:\: \myodot \equiv 3 \,,\qquad
    \text{(2+2)}:\:   \myodot \equiv 4 \,.
\end{equation}
Then the solar mass-splitting can be written for all schemes as
\begin{equation}
    \dms = m^2_\myodot - m_1^2 > 0\,.
\end{equation}
One advantage of the labeling introduced above is that we can use the
parameter $\Delta m^2_{41}\equiv m_4^2 - m_1^2$ to relate the
different schemes in a continuous way. The values of $\Delta m^2_{41}$
which correspond to the three schemes are given by
\begin{equation} \label{eq:dm41table}
    \begin{aligned}
	\text{(3+1)}_a &:\: & \Delta m^2_{41} &= \dml + \dms \,, \\
	\text{(3+1)}_b &:\: & \Delta m^2_{41} &= \dml + \dms - \dma \,, \\
	\text{(2+2)}_A &:\: & \Delta m^2_{41} &= \dms \,.
    \end{aligned}
\end{equation}

It will be useful to factorize the mixing matrix $U$ into two
matrices: $U = O^{(2)} O^{(1)}$. Neglecting the complex phases in $U$
we write the matrices $O^{(i)}$ as a product of rotation matrices
$R_{ij}$ in the $(i,j)$ subspace with the angle $\theta_{ij}$. We
define
{ \catcode`*=\active \def*{\hspace{1pt}}
  \setlength{\arraycolsep}{2\arraycolsep}
  \begin{align}
      \label{eq:O1}
      O^{(1)} &= R_{14} R_{13} R_{12} =
      \begin{pmatrix}
	  c_{14}*c_{13}*c_{12}
	  & c_{14}*c_{13}*s_{12}
	  & c_{14}*s_{13}
	  & s_{14}
	  \\
	  -s_{12}
	  & c_{12}
	  & 0
	  & 0
	  \\
	  -s_{13}*c_{12}
	  & -s_{13}*s_{12}
	  & c_{13}
	  & 0
	  \\
	  -s_{14}*c_{13}*c_{12}
	  & -s_{14}*c_{13}*s_{12}
	  & -s_{14}*s_{13}
	  & c_{14}
      \end{pmatrix},
      \\[3mm]
      \label{eq:O2}
      O^{(2)} &= R_{34} R_{24} R_{23} =
      \begin{pmatrix}
	  1
	  & 0
	  & 0
	  & 0
	  \\
	  0
	  & c_{24}*c_{23}
	  & c_{24}*s_{23}
	  & s_{24}
	  \\
	  0
	  & -s_{34}*s_{24}*c_{23} - c_{34}*s_{23}
	  & -s_{34}*s_{24}*s_{23} + c_{34}*c_{23}
	  &  s_{34}*c_{24}
	  \\
	  0
	  & -c_{34}*s_{24}*c_{23} + s_{34}*s_{23}
	  & -c_{34}*s_{24}*s_{23} - s_{34}*c_{23}
	  &  c_{34}*c_{24}
      \end{pmatrix}
  \end{align}
  }
and order the flavor eigenstates in such a way that if all angles are
zero we have the correspondence $(\nu_e, \nu_\mu, \nu_\tau, \nu_s) =
(\nu_1, \nu_2, \nu_3, \nu_4)$.  In the following sections we will
consider the mixing parameters relevant in the three different classes
of experiments (SBL, solar and atmospheric) in more detail.

\section{SBL experiments}
\label{sec:SBL}

\subsection{SBL parameters}

In SBL experiments it is a good approximation to set the solar and
atmospheric mass-splittings to zero. Obviously, under this assumption
the two schemes (3+1)$_a$ and (3+1)$_b$ become equivalent.
Let us define the parameters $d_\alpha$ ($\alpha=e,\mu,\tau,s$) and
$A_{\mu;e}$ for the two schemes as
\begin{equation} \label{eq:defd}
    \begin{aligned}
	\text{(3+1)} &:\: 
	& d_\alpha &= |U_{\alpha 4}|^2 \,, \quad
	& A_{\mu;e} &= 4\,|U_{e4}|^2 |U_{\mu 4}|^2 \,, \\
	\text{(2+2)} &:\: 
	& d_\alpha &= |U_{\alpha 1}|^2 + |U_{\alpha 4}|^2 \,, \quad
	& A_{\mu;e} &= 4\,\left| U_{e1} U_{\mu 1}^*
	+ U_{e4} U_{\mu 4}^*  \right|^2 \,.
    \end{aligned}
\end{equation}
Then for both schemes the probability of SBL $\pnu{\mu} \to \pnu{e}$
transitions relevant for the accelerator experiments LSND, KARMEN and
NOMAD is given by
\begin{equation} \label{eq:app}
    P_{\nu_\mu\to\nu_e} = P_{\bar\nu_\mu\to\bar\nu_e}
    = A_{\mu;e} \; \sin^2 \frac{\dml L}{4E} \,,
\end{equation}
and the survival probabilities relevant in the SBL disappearance
experiments Bugey and CDHS are given by
\begin{equation} \label{eq:disapp}
    P_{\nu_\alpha\to\nu_\alpha} =
    P_{\bar\nu_\alpha\to\bar\nu_\alpha} =
    1 - 4\, d_\alpha (1-d_\alpha) \sin^2 \frac{\dml L}{4E} \,,
\end{equation}
where $\alpha = e$ refers to the Bugey and $\alpha = \mu$ to the CDHS
experiment. Here $L$ is the distance between source and detector and
$E$ is the neutrino energy. It is straightforward to see that in the
(3+1) scheme the relation
\begin{equation}\label{eq:3+1rel}
    \text{(3+1)}:\quad A_{\mu;e} = 4\,d_e d_\mu
\end{equation}
holds. Hence, there are only two independent SBL mixing parameters in
this case. However, in the (2+2) scheme the situation is qualitatively
different and there is only the restriction
\begin{equation}\label{eq:2+2rel}
    \text{(2+2)}:\quad
    A_{\mu;e} \leq 4\:\mbox{min}\left[d_e d_\mu\,,\:
    (1-d_e)(1-d_\mu) \right]
\end{equation}
which follows from unitarity of $U$, and therefore there remain three
independent mixing parameters for SBL experiments in the (2+2) scheme.

Note that the probabilities Eqs.~\eqref{eq:app} and \eqref{eq:disapp}
have the same form as in the two-neutrino case~\cite{ptv92,BGG}. The
amplitude $A_{\mu;e}$ can therefore be identified with the LSND mixing
angle:
\begin{equation}\label{eq:lsnd-ang}
    A_{\mu;e} \equiv \sql \,,
\end{equation}
and for the disappearance parameters the identification $4\,d_e(1-d_e)
\leftrightarrow \sin^22\theta_\mathrm{Bugey}$ (and similar for CDHS)
    can be made.

\subsection{Constraints form Bugey and CHOOZ}
\label{sec:BugeyCHOOZ}

Let us consider the constraints from reactor $\bar\nu_e$ disappearance
experiments Bugey and CHOOZ.\footnote{Note that the Palo Verde reactor
  experiment~\cite{PaloV} obtains a bound comparable to CHOOZ.  As the
  exact value of this bound has very little impact on our analysis we
  include for simplicity only the result of CHOOZ.} To this purpose we
introduce the parameter
\begin{equation}\label{eq:etae}
    \eta_e \equiv |U_{e1}|^2 + |U_{e\myodot}|^2 \,,
\end{equation}
which describes the fraction of the electron neutrino in the ``solar
sector'' and is related to $d_e$ by
\begin{equation}\label{eq:etaeRel}
    \begin{aligned}
	\text{(3+1)} &:\quad \eta_e + d_e \leq 1 \,, \\
	\text{(2+2)} &:\quad \eta_e = d_e \,.
    \end{aligned}
\end{equation}
The requirement that the electron neutrino must participate in
oscillations with $\dms$ in order to explain the solar neutrino
anomaly leads to $\eta_e \sim 1$. The result of the Bugey
experiment~\cite{bugey} constrains the combination $4\,d_e(1-d_e)$ to
be very small. Taking into account Eq.~\eqref{eq:etaeRel} and $\eta_e
\sim 1$ one obtains~\cite{BGG,GS}
\begin{equation}\label{eq:deBound}
    \left.\begin{aligned}
	\text{(3+1)} &:\: & d_e  \\
	\text{(2+2)} &:\: & 1-d_e
    \end{aligned} \right\}
    \lesssim 2\times 10^{-2} \quad\text{at 90\% \CL}
\end{equation}
in the relevant range of $\dml$.

The disappearance probability in the CHOOZ~\cite{CHOOZ}
experiment\footnote{Due to its value of ($E/L$) CHOOZ is sensitive to
  oscillations with $\dma$ rather than with $\dml$, and therefore it is
  considered as a long-baseline experiment. However, it turns out to be
  convenient to treat it together with the SBL experiments in the
  analysis.} can be written as
\begin{equation}\label{eq:choozProb}
    P_\Chooz = 1 - 2\, d_e(1-d_e) -
    A_\Chooz \, \sin^2 \frac{\dma L}{4E} \,,
\end{equation}
with
\begin{equation}\label{eq:ACH}
    \begin{aligned}
	\text{(3+1)} &:\: & A_\Chooz
	&= 4\,\eta_e(1-d_e-\eta_e) \,,\\
	\text{(2+2)} &:\: & A_\Chooz
	&= 4\,|U_{e2}|^2 |U_{e3}|^2 \,.
    \end{aligned}
\end{equation}
We use the result of this experiment in two ways. First, we constrain
the SBL parameter $d_e$ similar to Bugey as described in
Ref.~\onlinecite{GS}. Second, also the parameter $A_\Chooz$ is
constrained to small values. Comparing Eqs.~\eqref{eq:deBound} and
\eqref{eq:ACH} and noting that for the (2+2) schemes
$(1-d_e)=|U_{e2}|^2 + |U_{e3}|^2$ one can see that $A_\Chooz$ is very
small in this case because of the bound on $(1-d_e)$ implied by Bugey.
However, for the (3+1) schemes the additional information of CHOOZ is
important.  Taking again into account that the electron neutrino must
have significant mixing with $\nu_1$ and $\nu_\myodot$ to obtain solar
neutrino oscillations ($\eta_e\sim 1$), we obtain the bound
\begin{equation}\label{eq:choozBound}
    \text{(3+1)}:\quad
    1-d_e-\eta_e
    \lesssim 4 \times 10^{-2} \qquad\text{at}\quad
    90\%\: \mathrm{\CL}
\end{equation}
for the values of $\dma$ preferred by atmospheric neutrino
experiments.

To summarize, if oscillations of $\nu_e$ with $\dms$ are required,
bounds from reactor experiments imply in both types of mass schemes
that $\eta_e$ has to be close to 1: for (2+2) Eqs.~\eqref{eq:etaeRel}
and \eqref{eq:deBound} imply that ($1-\eta_e$) is bounded by Bugey,
whereas for (3+1) we obtain a somewhat weaker bound resulting from a
combination of the bounds from Bugey Eq.~\eqref{eq:deBound} and CHOOZ
Eq.~\eqref{eq:choozBound}:
\begin{equation}\label{eq:etaeBound}
    \begin{aligned}
	\text{(3+1)} &:\: & 1-\eta_e & \lesssim 6\times 10^{-2} \,, \\
	\text{(2+2)} &:\: & 1-\eta_e & \lesssim 2\times 10^{-2} \,.
    \end{aligned}
\end{equation}

These bounds can also be translated into bounds on the mixing angles
contained in the $O^{(1)}$ factor of the leptonic mixing matrix. One
of three angles in the matrix $O^{(1)}$ is the solar angle
$\theta_{1\myodot}$, which has to be
large~\cite{Gonzalez-Garcia:2000aj} in order to account for the
results of solar neutrino oscillation
experiments~\cite{sksol,chlorine,sage,gallex_gno,sno}.  Then the
bounds given in Eq.~\eqref{eq:etaeBound} imply in all mass schemes
that the other two angles have to be small.

\subsection{Data used from SBL experiments}
\label{sec:SBLdata}

In this section we describe the experimental data from SBL experiments
which we are using for our statistical analysis. We divide the
$\chi^2$-function describing the SBL experiments into two parts:
\begin{equation}\label{eq:chi2SBL}
    \begin{aligned}
	\chi^2_\Sbl&(\dml,\thl,d_e,d_\mu)= \\
	&\chi^2_\Nev(\dml,\thl,d_e,d_\mu) +
	\Delta\chi^2_\Lsnd(\dml, \thl)\,.
    \end{aligned}
\end{equation}
Here $\chi^2_\Nev$ contains the information from the experiments
Bugey, CDHS, KARMEN, NOMAD and CHOOZ, which find {\it no evidence}
(NEV) for neutrino oscillations, while $\chi^2_\Lsnd$ includes the
information of LSND, which is the only SBL experiment reporting an
evidence for oscillations. For what concerns the parameter dependence
shown in Eq.~\eqref{eq:chi2SBL} one has to keep in mind that in the
(3+1) scheme $\thl$ is related to $d_e$ and $d_\mu$ via
Eqs.~\eqref{eq:3+1rel} and \eqref{eq:lsnd-ang}, but in the context of
(2+2) schemes all these three parameters are independent.

The Bugey experiment~\cite{bugey} searches for $\bar\nu_e$
disappearance at the distances 15~m, 40~m and 95~m away from a nuclear
reactor. As input data for our analysis we use Fig.~17 of
Ref.~\onlinecite{bugey}, where the ratios of the observed events to
the number of expected events in case of no oscillations are shown in
25 bins in positron energy for the positions 15~m and 40~m, and 10
bins for the position 95~m.  The CDHS experiment~\cite{CDHS} searches
for $\nu_\mu$ disappearance by comparing the number of events in the
so-called back and front detectors at the distances $L_{\mathrm{back}}
= 885$ m and $L_\mathrm{front} = 130$~m, respectively, from the
neutrino source. The data is given in Tab.~1 of Ref.~\onlinecite{CDHS}
as ratios of these event numbers in 15 bins of ``projected range in
iron''. The KARMEN experiment~\cite{KARMEN} looks for $\bar\nu_e$
appearance in a $\bar\nu_\mu$ beam. We use the number of positron
events in 9 bins of positron energy as given in Fig.~2(b) of the
second reference in~\cite{KARMEN}.  Our re-analysis of the experiments
Bugey, CDHS and KARMEN is described in detail in Ref.~\onlinecite{GS}.
To include the results on the $\nu_\mu\to\nu_e$ appearance channel
obtained by the NOMAD experiment~\cite{nomad} we perform an analysis
similar to the one of KARMEN. We use the 14 data points of the energy
spectrum of $\nu_e$ charged current events given in Fig.~2 of the
first reference in~\cite{nomad}.

We include the result of the CHOOZ experiment~\cite{CHOOZ} by means of
the $\chi^2$-function
\begin{equation}
    \chi^2_\Chooz =
    \frac{(\langle P_\Chooz \rangle - P_\mathrm{exp} )^2}
    {\sigma_\mathrm{stat}^2 + \sigma_\mathrm{syst}^2} \,,
\end{equation}
where $P_\mathrm{exp} = 1.01$, $\sigma_\mathrm{stat} = 2.8\%$,
$\sigma_\mathrm{syst} = 2.7\%$~\cite{CHOOZ} and $P_\Chooz$ is given in
Eq.~\eqref{eq:choozProb}. In the (2+2) case we adopt the approximation
$A_\Chooz = 0$ (see Sec.~\ref{sec:BugeyCHOOZ}) and hence
$\chi^2_\Chooz$ depends only on the parameter $d_e$. For the (3+1)
case $\chi^2_\Chooz$ depends on the two independent parameters $d_e$
and $\eta_e$. Apart from the requirement $\eta_e\sim 1$ we are not
interested in the exact value of this parameter and we will always
minimize with respect to it. In our approximation the CHOOZ experiment
is the only one sensitive to the small value $(1-\eta_e)$ (see
following sections), and therefore the minimization with respect to
$\eta_e$ is trivial and yields $A_\Chooz=0$. Again we are using only
the information on $d_e$ from CHOOZ in Eq.~\eqref{eq:chi2SBL}, which
is independent of $\dma$.  Therefore, the dependence on $\eta_e$ is
not shown in Eq.~\eqref{eq:chi2SBL}.

The total number of data points for all NEV experiments is
\begin{equation} \label{eq:NNEV}
    N_\Nev = 60_\mathrm{(Bugey)} + 15_\mathrm{(CDHS)} +
    9_\mathrm{(KARMEN)}+ 14_\mathrm{(NOMAD)} + 1_\mathrm{(CHOOZ)} = 99 \,.
\end{equation}

To include the detailed structure of the LSND
experiment~\cite{LSND,LSND2001} the LSND collaboration~\cite{LSNDpriv}
has provided us with a table of the likelihood function obtained in
the final analysis of their data~\cite{LSND2001} as a function of the
two-neutrino parameters $\dml$ and $\sql$. Contours of this likelihood
function corresponding to 90\% and 99\% \CL\ are shown in Fig.~27 of
Ref.~\onlinecite{LSND2001}.  The reason that we can use this
likelihood function, which was obtained in a two-neutrino analysis,
also in the four-neutrino case is that the relevant four-neutrino
probability Eq.~\eqref{eq:app} has the same form as the two-neutrino
probability. We include the LSND likelihood function in our analysis
by transforming it into a $\chi^2$-function according to~\cite{pdg}
$\chi^2 = \mathrm{const} - 2\, \ln \mathcal{L}$.  Because of the
event-by-event based likelihood analysis performed by the LSND
collaboration we cannot use any information on the absolute value of
the $\chi^2$-function. Therefore, as indicated already in
Eq.~\eqref{eq:chi2SBL}, we use in our analysis only the $\Delta\chi^2$
relative to its minimum:
\begin{equation}\label{eq:chi2LSND}
    \Delta\chi^2_\Lsnd(\dml,\thl) \equiv
    2\,\ln \mathcal{L}^\mathrm{max}_\Lsnd -
    2\,\ln \mathcal{L}_\Lsnd(\dml,\thl)\,.
\end{equation}
In this way we are able to include the LSND data in an optimal way, as
we are using directly the analysis performed by the experimental
group.

\section{Solar neutrino experiments}
\label{sec:solar}

For solar neutrino oscillations it is a good approximation to work in
the limit $\dml \to \infty$ and $\dma \to \infty$, so that
oscillations induced by the LSND and atmospheric mass-squared
differences are completely averaged out.  Moreover, in
Refs.~\onlinecite{Concha:2001uy,Concha:2001zi,Dooling:2000sg,Giunti:2000wt}
solar neutrino oscillations in (2+2) schemes has been studied using
the approximation $\eta_e=1$, which is justified by the Bugey bound
Eq.~\eqref{eq:deBound}. The results obtained there can be applied also
to (3+1) mass schemes, if again $\eta_e=1$ is adopted.  Note however,
that in this case only the somewhat weaker bound shown in
Eq.~\eqref{eq:etaeBound} applies. The solar oscillation probabilities
obtained in these works are valid up to terms of order $(1-d_e)^2$ for
(2+2) and $[d_e^2,~(1-\eta_e)^2]$ for (3+1).  Setting $\eta_e=1$
reduces the matrix $O^{(1)}$ in all cases to $R_{1\myodot}$,
eliminating the other two mixing angles.

Under these approximations solar neutrino oscillations do not
distinguish between (3+1) and (2+2) schemes, and depend only on the
three parameters $\dms$, $\theta_\Sol$ and
$\eta_s$~\cite{Dooling:2000sg}. The solar mixing angle $\theta_\Sol =
\theta_{1\myodot}$ is given by
\begin{equation}
    \tan^2\theta_\Sol \equiv \frac{|U_{e\myodot}|^2}{|U_{e1}|^2}
\end{equation}
and corresponds to $\theta_{12}$ in the notation of
Refs.~\onlinecite{Concha:2001uy,Concha:2001zi}; it can be taken in the
interval $0\leq\theta_\Sol\leq\pi/2$ without loss of generality.  The
parameter $\eta_s$ is defined by
\begin{equation}\label{eq:defetas}
    \eta_s \equiv |U_{s1}|^2 + |U_{s\myodot}|^2
\end{equation}
and corresponds to $c_{23}^2 c_{24}^2$ in the notation of
Refs.~\onlinecite{Concha:2001uy,Concha:2001zi}. This parameter
describes the fraction of the sterile neutrino participating in solar
neutrino oscillations: for $\eta_s = 0$ solar electron neutrinos
oscillate only into active neutrinos, whereas $\eta_s = 1$ corresponds
to pure $\nu_e \to \nu_s$ oscillations.\footnote{The parameter
  $\eta_s$ is similar to the parameters $A$ and $c_s$, which have been
  introduced in Refs.~\onlinecite{OY} and \onlinecite{BGGSbbn},
  respectively, to describe the effect of (2+2) mass schemes in Big-Bang
  nucleosynthesis.} Thus this \emph{mixing-type} parameter can be
interpreted as a \emph{model} parameter interpolating between the
approximate forms for the leptonic mixing matrix given in
Refs.~\onlinecite{ptv92} and \onlinecite{pv93}, respectively.

To include the information from solar neutrino experiments in our
analysis we make use of the results obtained in the four-neutrino
analysis performed in Ref.~\onlinecite{Concha:2001zi}. The
experimental data used in this work is the solar neutrino rate of the
chlorine experiment Homestake~\cite{chlorine}, the weighted average
rate of the gallium experiments SAGE~\cite{sage}, GALLEX and
GNO~\cite{gallex_gno}, as well as the 1258-day Super-Kamiokande data
sample~\cite{sksol} in form of the recoil electron energy spectrum for
both day and night periods, each of them given in 19 data bins, and
the recent result from the charged current event rate at
SNO~\cite{sno}. The total number of data points contained in
$\chi^2_\Sol$ is
\begin{equation}\label{eq:Nsol}
    N_\Sol=3_\mathrm{(Cl,Ga,SNO)}+ 38_\mathrm{(SK)} = 41 \,.
\end{equation}
Details of the solar neutrino analysis can be found in
Refs.~\onlinecite{Concha:2001zi,bahcall,GMPV} and references therein.

\begin{figure}[t] \centering
    \includegraphics[width=0.6\linewidth]{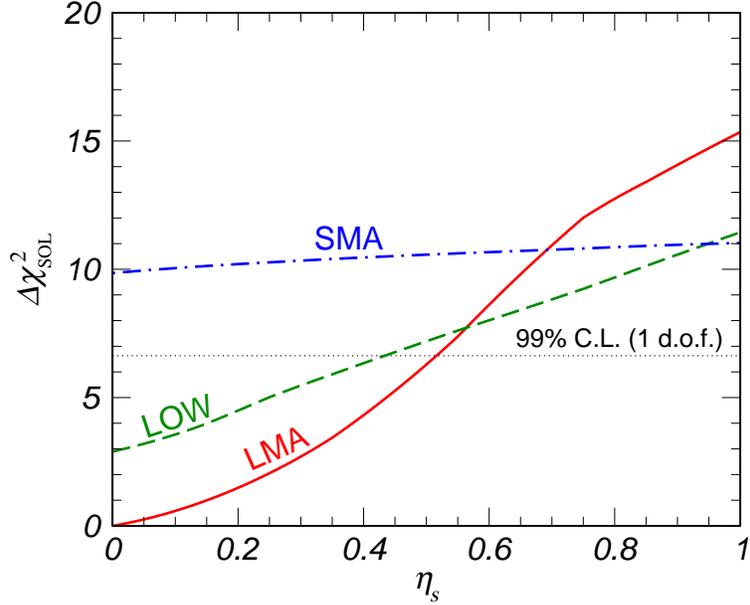}
    \caption{\label{fig:etas_sol}%
      $\Delta\chi^2_\Sol$ as a function of $\eta_s$ for the different
      solutions to the solar neutrino problem, as presented in Fig.~3
      of Ref.~\onlinecite{Concha:2001zi}.}
\end{figure}

To include the results of Ref.~\onlinecite{Concha:2001zi} in our
analysis we use $\chi^2_\Sol$ as a function of $\eta_s$ (minimized
with respect to the other two parameters $\dms$ and $\theta_\Sol$)
shown in Fig.~3 of Ref.~\onlinecite{Concha:2001zi}, which we reproduce
in Fig.~\ref{fig:etas_sol}. The $\chi^2$ is shown relative to the
global minimum, which lies in the large mixing angle (LMA) region and
has the value $(\chi^2_\Sol)_\mathrm{min} = 35.3$ for $N_\Sol - 3 =
38$ degrees of freedom (\dof).  The three lines in the figure are
obtained by requiring that the solution of the solar neutrino problem
lies in the three regions LMA, low/quasi-vacuum (LOW) and small mixing
angle (SMA), respectively. Note that $\chi^2_\Sol(\eta_s)$ is the same
for all mass schemes. We clearly see from this figure that solar
neutrino data prefers $\eta_s=0$, {\it i.e.}\ pure active
oscillations. At 99\% \CL\ there is the upper bound from the solar
data~\cite{Concha:2001zi}:
\begin{equation}\label{eq:etasSol}
    \text{solar data:}\quad \eta_s \leq 0.52\,.
\end{equation}

\section{Atmospheric neutrino experiments}
\label{sec:atmospheric}

For the oscillations of atmospheric neutrinos it is a good
approximation to set $\dms$ to zero and to also assume the limit $\dml
\to \infty$. In
Refs.~\onlinecite{Concha:2001uy,Concha:2001zi,atm4nuFit} fits of
atmospheric neutrino data in a (2+2) framework have been performed by
making use of the Bugey constraint Eq.~\eqref{eq:deBound} and setting
$\eta_e = 1$. The approximations $\dms=0$ and $\eta_e = 1$ imply that
the electron neutrino decouples completely from atmospheric neutrino
oscillations. In (3+1) spectra the contribution of electron neutrinos
to atmospheric oscillations is limited by the somewhat weaker bound
shown in Eq.~\eqref{eq:etaeBound}; however, in Ref.~\onlinecite{GMPV}
it was found that a $\nu_e$ contamination small enough not to spoil
the result of the CHOOZ experiment has only a very small effect on the
quality of the fit of atmospheric neutrino data. Therefore, it is
justified to adopt the approximation $\eta_e=1$ also for (3+1)
schemes~\cite{Maltoni:2001mt}.

Under these assumptions, atmospheric neutrino oscillations do not
distinguish between (3+1) and (2+2) schemes, and reduce to an
effective three-neutrino problem involving only the flavors $\nu_\mu$,
$\nu_\tau$, $\nu_s$, the mass eigenstates $\nu_2$, $\nu_3$, $\nu_4$
and the mixing matrix $O^{(2)}$ defined in Eq.~\eqref{eq:O2}. The
$\chi^2_\Atm$ function depends on the four parameters $\dma$,
$\theta_{23}$, $\theta_{34}$ and
$\theta_{24}$~\cite{Concha:2001uy,Concha:2001zi}, and to cover the
full physical parameter space one can choose the ranges $0\leq
(\theta_{24},\: \theta_{34}) \leq \pi/2$ and $-\pi/2 \leq \theta_{23}
\leq \pi/2$.\footnote{Note that $(\theta_{34}, \theta_{24},
  \theta_{23})$ in our notation correspond to $(\vartheta_{24},
  \vartheta_{23}, \vartheta_{34})$, respectively, in the notation of
  Refs.~\onlinecite{Concha:2001uy,Concha:2001zi}.} Therefore, in
addition to the two parameters $\dma$ and $\theta_\Atm \equiv
\theta_{23}$ corresponding to the two-neutrino parameters, we need two
more angles to describe atmospheric neutrino oscillations in a
four-neutrino framework~\cite{Dooling:2000sg}.

To understand the physical meaning of the angles $\theta_{24}$ and
$\theta_{34}$ let us consider their relation to the parameters $d_\mu$
and $d_s$, which we have defined in Eq.~\eqref{eq:defd}. Under the
approximation $\eta_e = 1$, we obtain in all the schemes
\begin{equation}\label{eq:dmuds}
    \begin{aligned}
	d_\mu & = |O^{(2)}_{\mu 4}|^2 = s_{24}^2 \,,\\
	d_s   & = |O^{(2)}_{s4}|^2 = c_{24}^2 c_{34}^2 \,.
    \end{aligned}
\end{equation}
The quantity ($1-d_\mu$) [($1-d_s$)] corresponds to the fraction of
the muon [sterile] neutrino participating in ``atmospheric'' neutrino
oscillations.  For $d_\mu = s_{24}^2 = 0$ the muon neutrino lies
completely in the atmospheric sector, while for the (strongly
disfavored) case $d_\mu = 1$ there are no oscillations of $\nu_\mu$
with the scale $\dma$. Hence, atmospheric data will constrain $d_\mu$
to be small.  Depending on the value of $\dml$, the bound on $d_\mu$
is strengthened by the $\pnu{\mu}$ SBL disappearance experiment
CDHS~\cite{BGGS,Maltoni:2001mt}. Similarly, $d_s = c_{24}^2 c_{34}^2 =
1$ corresponds to pure active atmospheric oscillations, whereas for
$d_s = 0$ the sterile neutrino fully participates in oscillations with
$\dma$.
The cases correspond to the approximations used in the early
papers~\cite{ptv92,pv93}: the \emph{mixing-type} parameter $d_s$ can
be interpreted as a \emph{model} parameter interpolating between the
approximate forms for the leptonic mixing matrix given in
Refs.~\onlinecite{ptv92} ($d_s=0$) and \onlinecite{pv93} ($d_s=1$),
respectively.

For the atmospheric data analysis we use the following data from the
Super-Kamiokande experiment~\cite{skatm}: $e$-like and $\mu$-like data
samples of sub- and multi-GeV, each given as a five-bin zenith-angle
distribution, up-going muon data including the stopping (5 bins in
zenith angle) and through-going (10 angular bins) muon fluxes.
Further, we use the recent update of the MACRO~\cite{macroNew}
up-going muon sample (10 angular bins). We obtain a total number of
data points contained in $\chi^2_\Atm$ of
\begin{equation}\label{eq:Natm}
    N_\Atm=35_\mathrm{(SK)}+ 10_\mathrm{(MACRO)} = 45 \,.
\end{equation}
For further details of the atmospheric neutrino analysis see
Refs.~\onlinecite{Concha:2001uy,Concha:2001zi,GMPV} and references
therein.

\begin{figure}[t] \centering
    \includegraphics[width=0.7\linewidth]{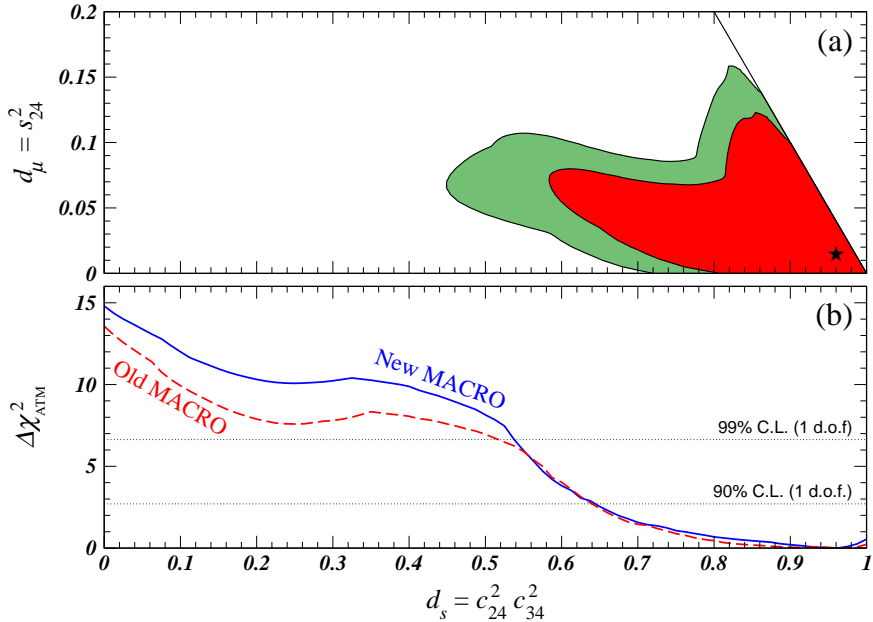}
    \caption{\label{fig:atmFig}%
      (a) 90\% and 99\% \CL\ allowed regions for the parameters $d_s =
      c_{24}^2 c_{34}^2$ (ordinate) and $d_\mu = s_{24}^2$ (abscissa)
      from atmospheric neutrino data. The best fit point is marked
      with a star. (b) $\Delta\chi^2_\Atm$ as a function of $d_s$
      using old~\cite{macroOld} and new~\cite{macroNew} MACRO data.
      Also shown are the $\Delta\chi^2$-values corresponding to 90\%
      and 99\% \CL\ for 1 \dof.}
\end{figure}

In Fig.~\ref{fig:atmFig} we show the results of our atmospheric
neutrino analysis regarding the angles $\theta_{24}$ and
$\theta_{34}$. This figure corresponds to Fig.~6 of
Ref.~\onlinecite{Concha:2001zi}, but now using the updated results of
MACRO. In the upper panel we show the 90\% and 99\% \CL\ allowed
regions (2 \dof) for the parameters $d_s = c_{24}^2 c_{34}^2$ and
$d_\mu = s_{24}^2$.  To obtain these regions we minimize $\chi^2_\Atm$
with respect to the other two parameters $\theta_\Atm$ and $\dma$. As
expected, atmospheric data constrains $d_\mu$ to small values,
implying a large fraction of $\nu_\mu$ participating in atmospheric
oscillations. For what concerns the parameter $d_s$, values close to 1
are preferred, which means that $\nu_\mu$ oscillates mainly to active
neutrinos. This can be seen clearly from the lower panel of
Fig.~\ref{fig:atmFig}, where we display $\Delta\chi^2_\Atm(d_s) \equiv
\chi^2_\Atm(d_s) - (\chi^2_\Atm)_\mathrm{min}\,$. Here
$(\chi^2_\Atm)_\mathrm{min} = 27.9$ for $N_\Atm - 4 = 41$ \dof\ and
$\chi^2_\Atm$ is minimized with respect to all other parameters. We
show the line for the updated MACRO data~\cite{macroNew} and compare
with the line obtained from the old MACRO data~\cite{macroOld}, which
corresponds to the data used in Ref.~\onlinecite{Concha:2001zi}. For
large values of $d_s$ the lines are very similar, however for small
values the fit gets worse. This means that atmospheric data get
stronger in rejecting a sterile component in atmospheric neutrino
oscillations.

\section{Four-neutrino parameters in the combined analysis}
\label{sec:paramSummary}

\begin{table}[t] \centering
    \begin{tabular}{l|l}
	\hline\hline
	data set          & parameters \\ \hline
	solar             & $\dms\,$, $\theta_\Sol\,$, $\eta_s$ \\
	atmospheric       & $\dma\,$, $\theta_\Atm$,
	$\theta_{24}\,$, $\theta_{34}$ \\
	SBL appearance    & $\dml\,$, $\theta_\Lsnd$ \\
	SBL disappearance & $\dml\,$, $d_e\,$, $d_\mu$ \\ \hline\hline
    \end{tabular}
    \caption{\label{tab:param}%
      Four-neutrino parameters for the different data sets.}
\end{table}
In the previous sections we have discussed the parameterization of the
four-neutrino problem for the different data sets separately. We
summarize our choice of parameters in Tab.~\ref{tab:param}. Note that
for (3+1) schemes $\eta_e$ is an additional independent parameter, but
we do not list it in Tab.~\ref{tab:param} because in our approximation
CHOOZ is the only experiment sensitive to it and we always minimize
with respect to it. In (2+2) schemes we have $\eta_e=d_e$ according to
Eq.~\eqref{eq:etaeRel}. We have chosen the parameters listed in
Tab.~\ref{tab:param} in such a way that they have a well-defined
physical meaning in the context of a given data set. Note that this
physical interpretation is {\it independent} of the mass scheme: for
example, regardless of whether we assume (3+1) or (2+2) schemes,
$\eta_s$ is the fraction of sterile neutrinos in solar oscillations,
$\theta_{24}$ describes the fraction of $\nu_\mu$ in atmospheric
oscillations (see Eq.~\eqref{eq:dmuds}), and $\sql$ is the SBL
$\pnu{\mu}\to\pnu{e}$ amplitude, and so on. The fact that it is
possible to describe the results of any of the given set of
experiments in terms of physical quantities independent of the mass
scheme implies that none of the considered data sets (solar,
atmospheric, SBL appearance or SBL disappearance) can be used {\it on
its own} to distinguish between different mass spectra. This follows
from the approximation $\eta_e \approx 1$, which is motivated by the
bounds from reactor neutrino experiments, and from the strong
hierarchy among the mass-squared differences indicated by the
data. This hierarchy implies that for any set of experiments only one
mass scale is relevant. In the following we show in detail that the
differences between the mass schemes manifest themselves only if two
or more data sets are combined, {\it i.e.}\ if the relation among
parameters belonging to different data sets is considered.

\begin{figure}[t] \centering
    \includegraphics[width=0.9\linewidth]{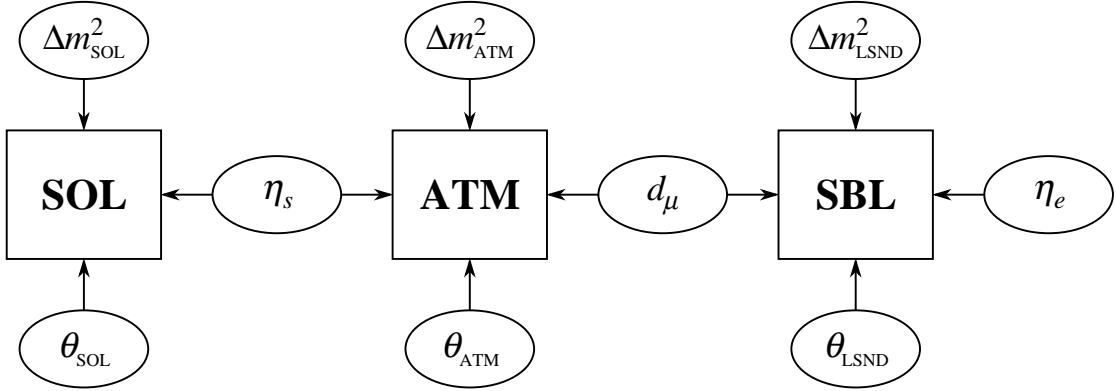}
    \caption{\label{fig:parameters}%
      Parameter dependence of the three data sets solar, atmospheric
      and SBL. Exact definitions of the parameters are given in
      Secs.~\ref{sec:SBL}, \ref{sec:solar} and \ref{sec:atmospheric}.}
\end{figure}

For the combined analysis we will describe neutrino oscillations by
means of the following parameters: beside the three mass-squared
differences $\dms$, $\dma$ and $\dml$ we use the six parameters
\begin{equation}\label{eq:param}
    \theta_\Sol\,, \: \theta_\Atm \,, \:
    \theta_\Lsnd\,, \:
    \eta_s\,, \: d_\mu \,,\: \eta_e \,.
\end{equation}
It is easy to check that indeed for all mass schemes these six
parameters --~defined as in the previous sections~-- can be used to
describe, in a physically more convenient way, the most general CP
conserving leptonic mixing matrix~\cite{Schechter:1980gr}. Each of the
$\chi^2$-functions describing the three data sets (SBL, solar,
atmospheric) depends only on a sub-set of these parameters:
\begin{equation}\label{eq:chi2param}
    \chi^2_\Sol(\dms, \theta_\Sol, \eta_s)\,,\quad
    \chi^2_\Atm(\dma, \theta_\Atm, d_\mu, \eta_s) \,,\quad
    \chi^2_\Sbl(\dml, \theta_\Lsnd, d_\mu, \eta_e) \,.
\end{equation}
We illustrate the parameter dependence of the data sets in
Fig.~\ref{fig:parameters}. The three angles $\theta_\Sol$,
$\theta_\Atm$, $\theta_\Lsnd$ are related directly to the amplitude of
the oscillations in the corresponding experiments solar, atmospheric
and LSND, while the two quantities $\eta_s$ and $d_\mu$ account for
the coupling between different data sets. As indicated, the parameter
$\eta_s$ is in common to solar and atmospheric neutrino oscillations;
if we express it in terms of the atmospheric angles we obtain the
different relations, depending on the mass schemes:
\begin{equation} \label{eq:etasAtm}
    \begin{aligned}
	\text{(3+1)}_a &:\: & \eta_s &= |O^{(2)}_{s2}|^2 =
	(s_{24} c_{34} c_\Atm -
	s_{34} s_\Atm)^2  \,, \\
	\text{(3+1)}_b &:\: & \eta_s &= |O^{(2)}_{s3}|^2 =
	(s_{24} c_{34} s_\Atm +
	s_{34} c_\Atm)^2 \,, \\
	\text{(2+2)}   &:\: & \eta_s &= |O^{(2)}_{s4}|^2 =
	c_{24}^2 c_{34}^2\,.
    \end{aligned}
\end{equation}
On the other hand the parameter $d_\mu$ is in common to SBL and
atmospheric neutrino oscillations; here the coupling is the same in
all mass schemes (see Eq.~\eqref{eq:dmuds}). Another important
difference between (3+1) and (2+2) arises due to the combination of
SBL appearance and disappearance experiments (see
Eqs.~\eqref{eq:3+1rel} and \eqref{eq:2+2rel}). There is no direct
coupling between solar and SBL oscillations, they do not depend on a
common parameter. This simple coupling of the data sets is a nice
feature of our parameterization (within the adopted approximations)
which renders the combined analysis possible, despite the large number
of parameters involved. Note that only the SBL experiments involve the
additional parameter $\eta_e$ explicitly, since --~as stated in
Secs.~\ref{sec:solar} and \ref{sec:atmospheric}~-- for what concerns
the analysis of the other two data sets it is safe to assume
$\eta_e=1$. In the (2+2) scheme we have the relation $\eta_e = d_e$
and this parameter is important for the SBL disappearance amplitude,
whereas in (3+1) only the long baseline reactor experiment CHOOZ is
sensitive to $\eta_e$ and we always minimize with respect to it.

For the unified analysis of all the mass schemes we will consider
$\dml$, $\dma$ and $\Delta m^2_{41}$ as the three independent
mass-squared differences. The $\chi^2$ as a function of $\Delta
m^2_{41}$ will have three local minima corresponding to the schemes
(3+1)$_a$, (3+1)$_b$ and (2+2)$_A$ at the values given in
Eq.~\eqref{eq:dm41table}. Beside this parameter indicating the scheme
we will display the results of our numerical analysis using the
following parameters. For the analysis of solar and atmospheric data
in Sec.~\ref{sec:sol+atm} we consider the $\chi^2$ as a function of
$\eta_s$; the results of the analysis of atmospheric and SBL data
(Sec.~\ref{sec:atm+SBL}) are given in the $(\dml, \sql)$ plane, while
for the fully global analysis (Sec.~\ref{sec:global}) we use all three
parameters $(\dml, \thl, \eta_s)$. Note that in a $\chi^2$ analysis
the size of the allowed regions depend crucially on the number of
parameters considered. Our aim was to identify which parameters
describe the most relevant features of the physics in each case.

Before closing this section let us note some subtleties related to our
parameterization. As described above, some of the parameters shown in
Eq.~\eqref{eq:param}, which we are using for our global fit, obey
different relations depending on the mass scheme considered. The
question arises of how to treat these different relations among
parameters in a common framework for the two mass schemes.  Indeed, we
are using the parameter $\Delta m^2_{41}$ to formally describe a
continuous transition between the vastly different mass spectra.

Let us consider a completely arbitrary parameterization of the general
four-neutrino problem~\cite{Schechter:1980gr}. We have 3 mass-squared
differences and 6 angles, {\it e.g.}\ the angles $\theta_{ij}$
introduced in Eqs.~\eqref{eq:O1} and \eqref{eq:O2}. Now one can think
of a fit to the data in this general parameterization. In practice it
is almost impossible to perform this general nine-parameter fit with
current computer technology. However, the results of such an analysis
would be six well separated regions in the nine-dimensional parameter
space, corresponding to the six mass schemes shown in
Fig.~\ref{fig:4spectra}. In these islands our approximations hold and
it makes sense to adopt a parameterization motivated by phenomenology.
The allowed regions for the parameters $\theta_{ij}$ can be mapped to
allowed regions for the parameters shown in Eq.~\eqref{eq:param},
which have a simple physical interpretation. Inside any given island
it is clear which relations among the new parameters have to be
applied. Obviously this reasoning is only valid under the assumption
that various regions are well separated. This assumption can be
justified by noting that to move continuously from a (2+2) to a (3+1)
scheme one has to break up the hierarchy among the mass-squared
differences, and we can expect that at least one data set will give a
large $\chi^2$ which separates the corresponding allowed regions.

\section{Analysis of solar and atmospheric neutrino data}
\label{sec:sol+atm}

In this section we combine solar and atmospheric neutrino data. In
Ref.~\onlinecite{Concha:2001zi} these data have been considered in the
(2+2) scheme. Here we discuss some slight changes in this case (due to
the updated MACRO data), we extend the analysis also to the case of
the (3+1) mass scheme in a way that allows a direct
comparison of the fit for these two schemes.

\begin{figure}[t] \centering
    \includegraphics[width=0.6\linewidth]{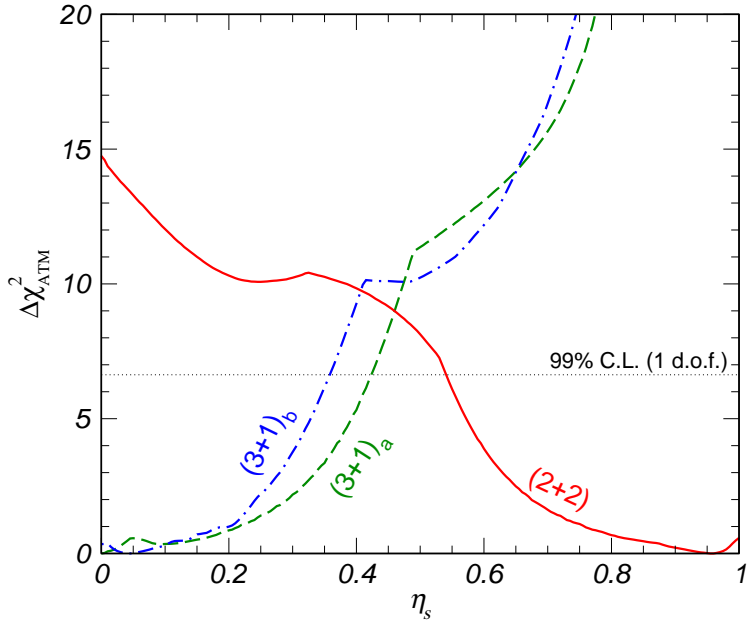}
    \caption{\label{fig:atm_etas}%
      $\Delta\chi^2_\Atm$ as a function of the fraction of the
      sterile neutrino in solar oscillations $\eta_s$ for all
      four-neutrino mass schemes, (3+1)$_a$, (3+1)$_b$ and (2+2).}
\end{figure}

Before combining the two data sets let us consider the impact of
atmospheric data alone on the parameter $\eta_s$, describing the
fraction of sterile neutrinos in solar oscillations. The relation of
$\eta_s$ to the atmospheric parameters is given in
Eq.~\eqref{eq:etasAtm} for the three mass schemes (3+1)$_a$, (3+1)$_b$
and (2+2). In Fig.~\ref{fig:atm_etas} we show
$\Delta\chi^2_\Atm(\eta_s) \equiv \chi^2_\Atm(\eta_s) -
(\chi^2_\Atm)_\mathrm{min}$ for the three cases, minimizing with
respect to the other parameters $\dma,\theta_\Atm$ and $d_\mu$. The
line corresponding to the (2+2) scheme is identical to the one shown
in the lower panel of Fig.~\ref{fig:atmFig} because of the (2+2)
relation $\eta_s = d_s = c_{24}^2 c_{34}^2$ (see
Eq.~\eqref{eq:etasAtm}). Atmospheric data prefers large values of
$\eta_s$, which corresponds to active $\nu_\mu$ atmospheric neutrino
oscillations. From the figure we can read off the 99\% \CL\ bound
\begin{equation}\label{eq:etasAtm2+2}
    \text{atmospheric data:}\quad \eta_s \geq 0.54 \quad\text{for (2+2) schemes}
\end{equation}
which is in disagreement with the bound \eqref{eq:etasSol} from solar
data. On the other hand, concerning (3+1) schemes in
Ref.~\onlinecite{peres} the very interesting fact was noted that {\it
  atmospheric} data give a constraint on the fraction of the sterile
neutrino participating in {\it solar} oscillations. From
Eq.~\eqref{eq:dmuds} it follows that $|O_{s2}^{(2)}|^2
+|O_{s3}^{(2)}|^2 = 1 - d_s$ is the fraction of sterile neutrinos in
atmospheric oscillations which should be small according to the
data~\cite{sksterile}. Comparing this with Eq.~\eqref{eq:etasAtm} we
expect that for (3+1) schemes atmospheric data prefers small values of
$\eta_s$. Indeed, from Fig.~\ref{fig:atm_etas} we find the 99\% \CL\
bounds
\begin{equation} \label{eq:etasAtm3+1}
    \text{atmospheric data:}
    \begin{cases}
	\eta_s \leq 0.35 & \text{for (3+1)$_a$ schemes}, \\
	\eta_s \leq 0.42 & \text{for (3+1)$_b$ schemes}
    \end{cases}
\end{equation}
which are even stronger than the one from solar data
Eq.~\eqref{eq:etasSol}.

From Fig.~\ref{fig:atm_etas} one can see that, although there are
quantitative differences between the two schemes (3+1)$_a$ and
(3+1)$_b$, the qualitative behavior is similar. Looking at
Eq.~\eqref{eq:etasAtm} it is easy to see that the relation between
$\eta_s$ and the atmospheric angles in the (3+1)$_a$ scheme reduces to
the one in the (3+1)$_b$ scheme under the transformation $\theta_\Atm
\to \theta_\Atm - \pi/2$.  Such a transformation, when applied to the
atmospheric oscillation probabilities, is equivalent to changing the
sign of $\dma$, hence we have $\chi^2_\Atm(\dma, \theta_\Atm - \pi/2,
\theta_{24}, \theta_{34}) = \chi^2_\Atm(-\dma, \theta_\Atm,
\theta_{24}, \theta_{34})$. Therefore, the origin of the difference
between the two schemes (3+1)$_a$ and (3+1)$_b$ can be explained in
two different ways. If we require $\dma > 0$, then we end up with two
different relations between $\eta_s$ and the atmospheric angles
$\theta_\Atm$, $\theta_{24}$ and $\theta_{34}$, as we have done so
far. Alternatively, if the definition of the atmospheric angles is
adjusted so that their relation with $\eta_s$ is the same in (3+1)$_a$
and (3+1)$_b$ schemes, then it is no longer sufficient to restrict to
the case $\dma > 0$, and the case $\dma < 0$ should be investigated as
well. In the latter approach, it is clear that the difference arise
because of the presence of {\it matter effects} in atmospheric
neutrino oscillations, which are sensitive to the sign of $\dma$.
Since in this work we are mainly interested in the comparison of (2+2)
with (3+1) in general, from now on we will always minimize with
respect to (3+1)$_a$ and (3+1)$_b$ by choosing from the two
corresponding values of $\Delta m^2_{41}$ the one with the lower
$\chi^2$.

\begin{figure}[t] \centering
    \includegraphics[width=0.85\linewidth]{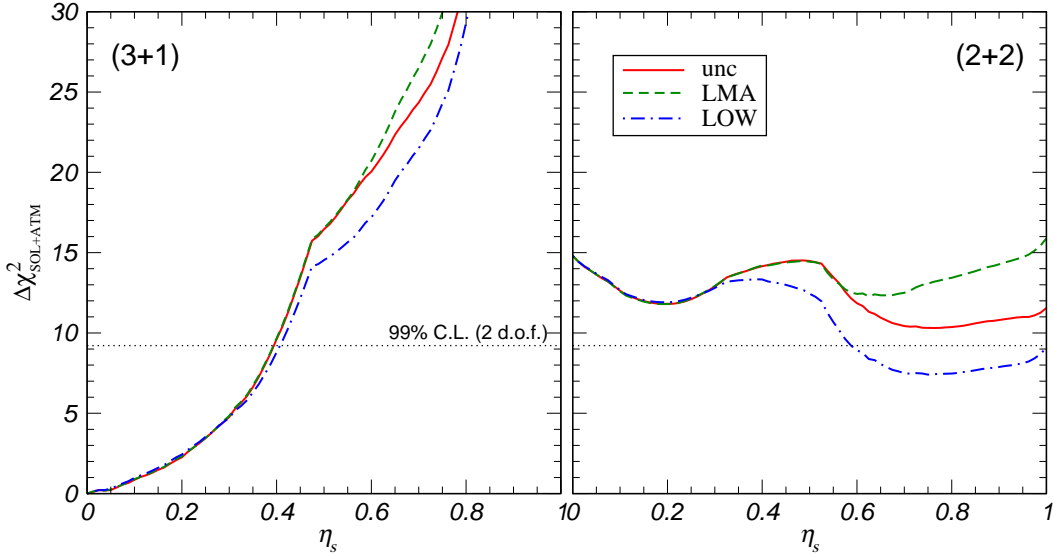}
    \caption{\label{fig:etas_sol+atm2}%
      Combined $\chi^2$-function for solar and atmospheric neutrino
      data for (3+1) and (2+2) for different solar neutrino solutions
      with respect to the global minimum (see text for details).}
\end{figure}

From Eqs.~\eqref{eq:etasSol}, \eqref{eq:etasAtm2+2} and
\eqref{eq:etasAtm3+1} one expects that combined solar and atmospheric
neutrino data will prefer (3+1) mass schemes over (2+2). In order to
quantify this statement let us consider the following
$\chi^2$-function:
\begin{equation}
    \chi^2_{\Sol+\Atm}(\eta_s,\Delta m^2_{41}) \equiv
    \chi^2_\Sol + \chi^2_\Atm \,
\end{equation}
where we minimize with respect to the parameters $\dms$, $\dma$,
$\theta_\Sol$, $\theta_\Atm$ and $d_\mu$. As explained before (see
Sec.~\ref{sec:parameters}) the parameter $\Delta m^2_{41}$ relates the
different schemes.  In Fig.~\ref{fig:etas_sol+atm2} we show the
$\Delta\chi^2$ projected onto the $\eta_s$-axis for the two regions of
$\Delta m^2_{41}$ corresponding to the schemes (3+1) and (2+2)
according to Eq.~\eqref{eq:dm41table}.  In {\it both} cases, (3+1) and
(2+2), we refer to the {\it same} minimum, which occurs for the (3+1)
scheme with $\eta_s=0$. For the dashed (dashed-dotted) line we
restrict the solar solution to be LMA (LOW), while for the solid line
(labeled ``unc'' in the figure) we choose the solution which gives the
weakest restriction (unconstrained). This corresponds to our current
knowledge of the solution to the solar neutrino problem.  The reason
why the line for LOW sometimes is below the one for the unconstrained case is
that it is referred to the minimum for the LOW solution, which is
higher than the minimum for LMA (which coincides with the minimum in
the unconstrained case) as can be seen from Fig.~\ref{fig:etas_sol}.
For the unconstrained case the minimum has the value
$(\chi^2_{\Sol+\Atm})_\mathrm{min} = 63.2$ for $N_\Sol + N_\Atm - 7 =
79$ \dof.

As expected, solar and atmospheric neutrino data prefer (3+1) because
in this case both can be explained by active neutrino oscillations.
The dotted line in Fig.~\ref{fig:etas_sol+atm2} shows the value of
$\Delta\chi^2=9.21$ corresponding to 99\% \CL\ for 2 \dof, which are
the two parameters $\eta_s$ and $\Delta m^2_{41}$. Therefore, for the
unconstrained and LMA cases (2+2) is disfavored at more than 99\% \CL\
with respect to (3+1). If we compare the local minimum with respect to
$\eta_s$ in (2+2) with the global minimum in (3+1) we find for the
unconstrained case
\begin{equation}\label{eq:dchi2_sol+atm}
    \Delta\chi^2 =
    \left( \chi^{2 \,\mathrm{(2+2)}}_{\Sol+\Atm}
    \right)_\mathrm{min} -
    \left( \chi^{2 \,\mathrm{(3+1)}}_{\Sol+\Atm}
    \right)_\mathrm{min}
    = 10.3 \,.
\end{equation}
Conversely, the LOW (2+2) solution is still allowed at the 99\% \CL.
The reason for this is, that the LOW solution is not as strong to
reject a sterile component in solar oscillations as the LMA solution
(see Fig.~\ref{fig:etas_sol}), so that the disagreement with
atmospheric data in the context of (2+2) schemes is somewhat weaker.
Let us note that for the unconstrained case large values of
$\eta_s\sim 0.76$ (corresponding to a large component of sterile
neutrino in solar oscillations) are slightly preferred over small
ones. This means that the inclusion of the updated MACRO results makes
atmospheric neutrino data slightly more powerful to reject the sterile
neutrino than the unconstrained solar data.

\section{Analysis of atmospheric and SBL neutrino data}
\label{sec:atm+SBL}

\begin{figure}[t] \centering
    \includegraphics[width=0.85\linewidth]{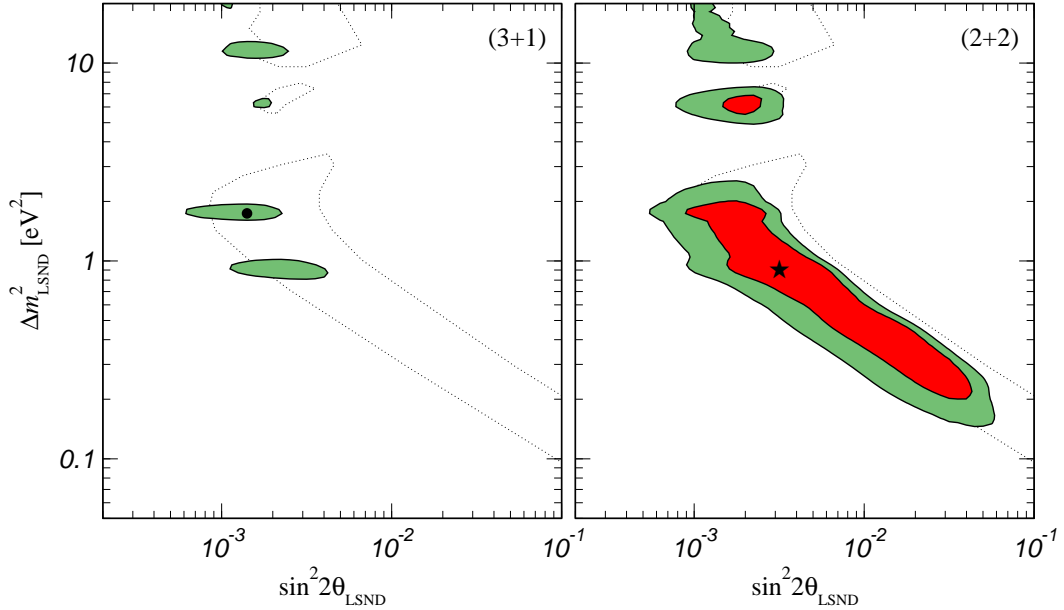}
    \caption{\label{fig:atm+SBL}%
      Combination of atmospheric and SBL data. We show projections of
      the three-dimensional 90\% and 99\% \CL\ regions corresponding
      to (3+1) and (2+2) in the ($\dml,\sql$) plane. The best fit
      point lies in (2+2) and is marked with a star, the local best
      fit point in (3+1) is marked with a circle. The doted line is
      the 99\% \CL\ region from LSND data alone~\cite{LSND2001}.}
\end{figure}

In this section we combine the data sets from atmospheric and SBL
neutrino experiments. To this purpose we consider the
$\chi^2$-function
\begin{equation}\label{eq:chi2atm+SBL}
    \chi^2_{\Atm+\Sbl} (\Delta m^2_{41},\dml,\thl) =
    \chi^2_\Atm + \chi^2_\Sbl\,.
\end{equation}
From Eq.~\eqref{eq:chi2param} we can see that the terms on the right
hand side in general depend on the parameters ($\dml,\dma,\theta_\Atm,
\thl, \eta_s, d_\mu, \eta_e$). To obtain the parameter dependence as
shown in Eq.~\eqref{eq:chi2atm+SBL} we proceed as follows. First, we
minimize $\chi^2_\Atm$ with respect to $\dma,\theta_\Atm$ and
$\eta_s$. In a second step we minimize with respect to $d_e$ and
$d_\mu$ by taking into account the relation~\eqref{eq:3+1rel} or
\eqref{eq:2+2rel}, depending on the mass scheme considered ({\it
i.e.}\ on the value of $\Delta m^2_{41}$).

The allowed regions in the parameter space $(\Delta m^2_{41}, \dml,
\thl)$ are given by $\Delta \chi^2 = 6.3$ ($11.3$) for 90\% (99\%)
\CL\ (3 \dof). In the right panel of Fig.~\ref{fig:atm+SBL} we show a
projection of the three-dimensional regions corresponding to the (2+2)
case, which include the best fit point ($\dml=0.91~\eVq$, $\sql =
3.16\times 10^{-3}$). One can see that the allowed regions cover a
large part of the two-neutrino allowed region by LSND alone, which is
shown as the dotted line. The allowed region disappears for values of
$\sql \gtrsim 0.06$ because of the constraint from the Bugey
experiment. At large values of $\dml$ the bounds from KARMEN and NOMAD
are important.  In the left panel we show the projection of the
three-dimensional volume corresponding to the (3+1) case with respect
to the global minimum, which lies in the (2+2) plane.  Only four small
islands appear at 99\% \CL. If we compare the local best fit point for
(3+1) at ($\dml = 1.74~\eVq$, $\sql = 1.41\times 10^{-3}$) with the
global best fit point we find
\begin{equation}\label{eq:dchi2_atm+SBL}
    \Delta\chi^2 =
    \left( \chi^{2 \,\mathrm{(3+1)}}_{\Atm+\Sbl}
    \right)_\mathrm{min} -
    \left( \chi^{2 \,\mathrm{(2+2)}}_{\Atm+\Sbl}
    \right)_\mathrm{min}
    = 6.9 \,.
\end{equation}
The conclusion from Fig.~\ref{fig:atm+SBL} and
Eq.~\eqref{eq:dchi2_atm+SBL} is that SBL data combined with
atmospheric data clearly prefer (2+2) over the (3+1) spectra.
Fig.~\ref{fig:atm+SBL} is a beautiful confirmation of the results of
our previous work~\cite{Maltoni:2001mt}, where we have analyzed a
similar data set in the (3+1) framework, but using a very different
statistical method. The reason for the (2+2) preference by the SBL
data is well known~\cite{peres,OY,barger98,BGGS,BGG,carlo} and can be
understood from Eqs.~\eqref{eq:3+1rel} and \eqref{eq:2+2rel}. For the
(3+1) case both $d_e$ and $d_\mu$ must be small because of Bugey and
CDHS, respectively, which leads to a double suppression of the LSND
amplitude $\sql$ according to Eq.~\eqref{eq:3+1rel}. In contrast, for
the (2+2) case $(1-d_e)$ and $d_\mu$ have to be small and
Eq.~\eqref{eq:2+2rel} implies only a linear suppression of $\sql$.

\section{Global analysis}
\label{sec:global}

\begin{figure}[t] \centering
    \includegraphics[width=0.85\linewidth]{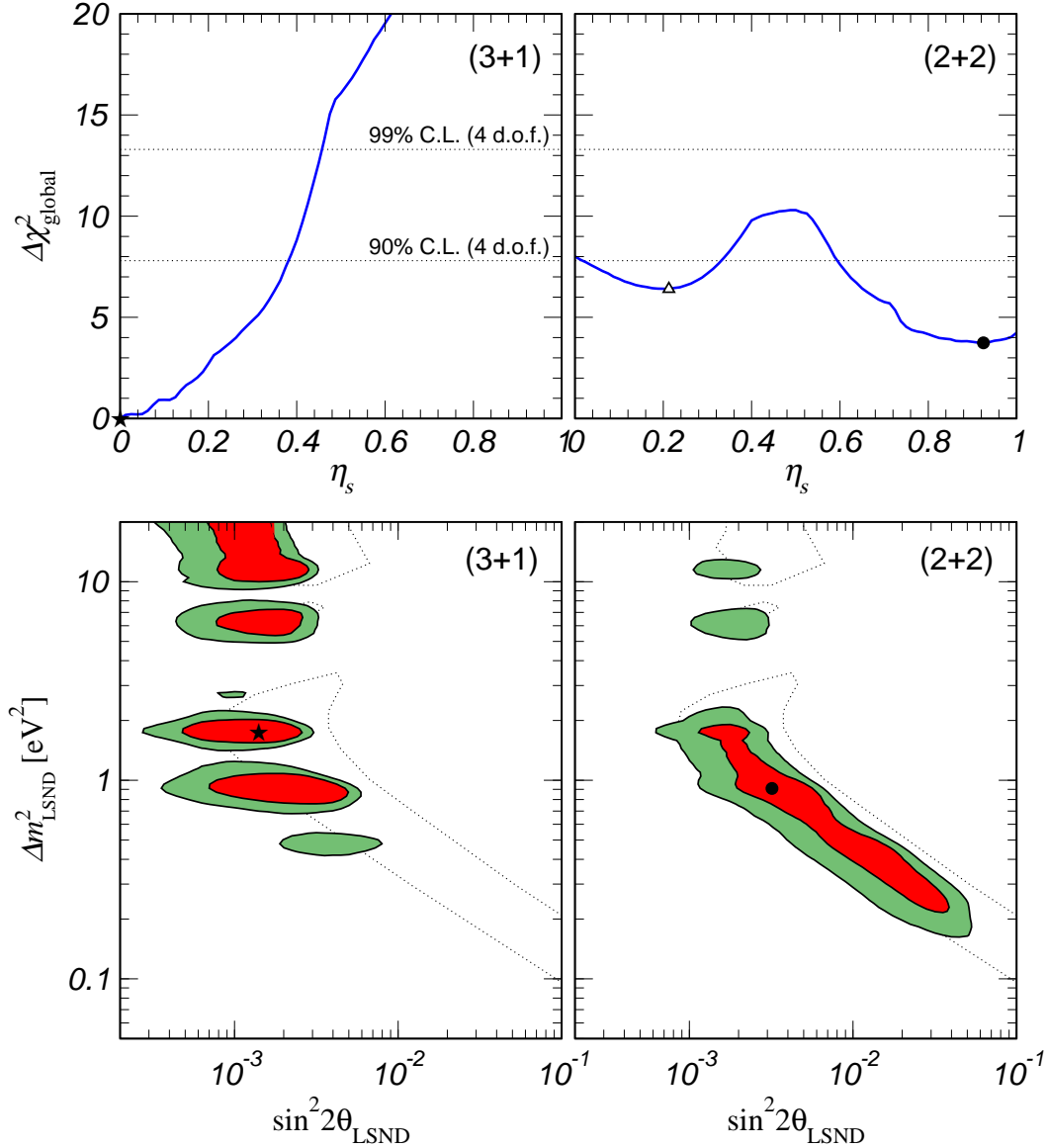}
    \caption{\label{fig:global}%
      Global combination of current neutrino oscillation data: solar,
      atmospheric and SBL. We show the $\Delta\chi^2_\mathrm{global}$
      as a function of $\eta_s$ (upper panels) and projections of the
      four-dimensional 90\% and 99\% \CL\ regions on the ($\dml,\sql$)
      plane (lower panels) (see text for details). The global best fit
      point lies in (3+1) and is marked with a star, the local best
      fit point in (2+2) is marked with a circle and the local minimum
      in (2+2) is marked with a triangle.  The doted line in the lower
      panels is the 99\% \CL\ region from LSND data
      alone~\cite{LSND2001}.}
\end{figure}

The results of the previous section, {\it i.e.}\ that atmospheric+SBL
data prefer (2+2) over (3+1), are in direct conflict with the results
of Sec.~\ref{sec:sol+atm}, where we have found that solar+atmospheric
data prefer (3+1) over (2+2). This shows that there is some tension in
the existing data in a four-neutrino framework, and to clarify the
situation it is necessary to perform a combined analysis of all the
data.  To this end we consider the $\chi^2$-function
\begin{equation}
    \chi^2_\mathrm{global}(\Delta m^2_{41}, \dml, \thl, \eta_s) =
    \chi^2_\Sol + \chi^2_\Atm + \chi^2_\Sbl \,.
\end{equation}
We minimize the right-hand side of this equation with respect to all
the parameters, except the ones shown on the left-hand side. Allowed
regions are given by
\begin{equation}
    \Delta\chi^2_\mathrm{global} =
    \chi^2_\mathrm{global} - (\chi^2_\mathrm{global})_\mathrm{min} =
    7.8 \, (13.3)
\end{equation}
for 90\% (99\%) \CL\ (4 \dof). From this equation we obtain two
four-dimensional volumes in the space of ($\Delta m^2_{41}, \dml,
\sql, \eta_s$) corresponding to (3+1) and (2+2), which we display in
Fig.~\ref{fig:global} in the following way. In the lower panels we
show projections of the four-dimensional volumes onto the
($\dml,\sql$) plane. In the upper panels we show
$\Delta\chi^2_\mathrm{global}$ minimized with respect to all
parameters except $\eta_s$. The projections of the four-dimensional
90\% and 99\% \CL\ volumes onto the $\eta_s$-axis are given by the
intersections of the solid lines in the upper panels with the
corresponding horizontal dotted lines.

\begin{table}[t] \centering
    \catcode`?=\active \def?{\hphantom{0}}
    \begin{tabular}{l|c@{\hspace{5mm}}c@{\hspace{5mm}}
	  c@{\hspace{5mm}}c@{\hspace{5mm}}c}
	\hline\hline
	& $\dml$ [$\eVq$] &  $\sql$ & $\eta_s$ & $d_\mu$ & $d_e$ \\
	\hline
	global best fit (3+1) & $1.74$ & $1.41\times 10^{-3}$ & $0.0?$
	& $1.98\times 10^{-2}$ & $1.79\times 10^{-2}$ \\
	best fit (2+2)        & $0.87$ & $3.55\times 10^{-3}$ & $0.93$
	& $6.56\times 10^{-3}$ & $0.99275$ \\
	local minimum (2+2)   & $0.87$ & $3.55\times 10^{-3}$ & $0.21$
	& $1.32\times 10^{-2}$ & $0.99275$ \\
	\hline\hline
    \end{tabular}
    \caption{\label{tab:globalbest}%
      Parameter values at the best fit points in (3+1) and (2+2) and
      at the local minimum in (2+2).}
\end{table}
Let us discuss the results of the global analysis. We find that the
global minimum lies in the (3+1) scheme. This minimum is marked as a
star in Fig.~\ref{fig:global}. In (2+2) there are two local minima:
the (2+2) best fit point is marked with a circle and corresponds to
large values of $\eta_s$, whereas the second local minimum (marked
with a triangle) occurs for small $\eta_s$.\footnote{Note that the two
  stars in the lower and upper panels actually correspond to the same
  single point in the four-dimensional space. The same holds for the two
  circles. We do not show the triangle in the lower right panel, because
  it would coincide with the circle (see Tab.~\ref{tab:globalbest}).}
The values of the parameters at these minima are given in
Tab.~\ref{tab:globalbest}. However, the difference between the best
fit points in (3+1) and (2+2) is not very big:
\begin{equation}\label{eq:dchi2_global}
    \Delta\chi^2 =
    \left( \chi^{2 \,\mathrm{(2+2)}}_\mathrm{global}
    \right)_\mathrm{min} -
    \left( \chi^{2 \,\mathrm{(3+1)}}_\mathrm{global}
    \right)_\mathrm{min}
    = 3.7 \,.
\end{equation}
We conclude that the schemes (3+1) and (2+2) give a comparable global
fit to the data. This can also be seen from the fact that there are
large allowed regions for both mass spectra.  The conflicting values
given in Eq.~\eqref{eq:dchi2_sol+atm} (for solar and atmospheric data)
and in Eq.~\eqref{eq:dchi2_atm+SBL} (for atmospheric and SBL data)
cancel each other to some extent. Solar plus atmospheric data seem to
be stronger than SBL data, therefore (3+1) is slightly preferred over
(2+2) in the global fit to current neutrino oscillation data.

The shape of the allowed regions in the ($\dml, \sql$) plane for (2+2)
(lower right panel of Fig.~\ref{fig:global}) is simliar to the one
expected from a two-neutrino analysis of SBL data alone.  In the
region $0.18~\eVq \lesssim \dml \lesssim 8~\eVq$ they follow closely
the two-neutrino LSND region. The slight shift to smaller values of
$\dml$ is because of the constraint from KARMEN.  Large values of
$\sql \gtrsim 0.06$ are excluded by Bugey and for $\dml \gtrsim
10~\eVq$ constraints from KARMEN and NOMAD are important.\footnote{All
the relevant SBL bounds are shown {\it e.g.}\ in Fig.~27 of
Ref.~\onlinecite{LSND2001}.  A combined analysis of LSND and KARMEN in
a two-neutrino framework has been performed in
Ref.~\onlinecite{eitel}.} In contrast, in the (3+1) case (lower left
panel of Fig.~\ref{fig:global}) the allowed regions consist of several
islands and are very different from the two-neutrino ones. The most
prominent islands are at the values $\dml \sim 0.9$, $1.7$, $6~\eVq$
and $\sql \sim 10^{-3}$. These are the values of $\dml$ where the
bounds of all NEV experiments have some marginal overlap with the LSND
allowed region~\cite{barger00,peres,carlo,GS,Maltoni:2001mt}.
However, at 99\% \CL\ appears an allowed region at $\dml \sim
0.5~\eVq$, and a very marginal island at $\dml \sim 2.5~\eVq$. There
is also an allowed region for large values of $\dml\gtrsim 10~\eVq$.
However, in this region there are further constraints from experiments
not included in our analysis, which are BNL E776~\cite{BNL}
($\pnu{\mu}\to\pnu{e}$ appearance) and CCFR~\cite{CCFR}
($\pnu{\mu}\to\pnu{e}$ appearance and $\pnu{\mu}$ disappearance).
Therefore, we do not display values of $\dml > 20~\eVq$.

Discussing the upper panels of Fig.~\ref{fig:global} we note that, as
already found in Sec.~\ref{sec:sol+atm}, large values of $\eta_s$
are preferred for the (2+2) case. Comparing the shape of the $\chi^2$
in the global analysis with the one shown in
Fig.~\ref{fig:etas_sol+atm2} we observe that the inclusion of SBL data
strengthen this trend to some extend. This implies a large component
of the sterile neutrino in solar neutrino oscillations and corresponds
to the LOW/quasi-vacuum solution of the solar neutrino problem (see
Fig.~\ref{fig:etas_sol}).  But also the local minimum for smaller
$\eta_s$ values, which corresponds to the LMA solar solution and
implies a large sterile component in atmospheric oscillations, is well
inside the 90\% \CL\ region. The difference in $\chi^2$ between the
two local minima in (2+2) is 2.7.  Moreover, the minima in $\chi^2$
are not very deep so that all values of $\eta_s$ between 0 and 1 are
within the 99\% \CL\ region.  Only values around 0.5 are excluded at
90 \% \CL. The results shown in Fig.~\ref{fig:global} were obtained by
using unconstrained solar data. We have also performed the analysis by
restricting solar data to the LMA and LOW region. The results are very
similar to the unconstrained case. For the LMA solution we obtain
approximately the solution corresponding to the local minimum in
(2+2), which means that (2+2) is sightly more disfavored against
(3+1), whereas for the LOW case the difference would become even
smaller than shown in Eq.~\eqref{eq:dchi2_global}.

Recently, solar neutrino data have been analyzed using a new
prediction of the $^8$B flux~\cite{bahcallNew}. From Tab.~3 of
Ref.~\onlinecite{bahcallNew} one can see that LMA becomes relatively
better than LOW. This would lead to an up-wards shift of the LOW line
in Fig.~\ref{fig:etas_sol} of approximately 3 units. Consequently
solar data becomes stronger in rejecting the sterile neutrino.
Regarding the four-neutrino analysis, this would disfavor the (2+2)
scheme slightly more against the (3+1) case.

In our framework it is also possible to test the fit of the (3+0)
scenario, where the solar and atmospheric neutrino problems are
explained by oscillations among three active neutrinos and the
explanation of LSND is left out. This would correspond to the Standard
Model situation. We obtain this case by considering the (3+1) scheme
(this fixes the parameter $\Delta m^2_{41}$) and setting the
parameters $d_e=d_\mu=\eta_s = 0$. Then the sterile neutrino is
completely decoupled and we are left with three active neutrinos and
the mass splittings $\dms$ and $\dma$. We find a difference in
$\chi^2$ to the best fit point of
\begin{equation}\label{eq:dchi2_3+0}
    \Delta\chi^2 =
    \chi^{2 \,\mathrm{(3+0)}}_\mathrm{global} -
    \left( \chi^{2 \,\mathrm{(3+1)}}_\mathrm{global}
    \right)_\mathrm{min}
    = 19.8 \,.
\end{equation}
For 4 \dof\ ($d_e,d_\mu,\eta_s,\Delta m^2_{41}$) this corresponds to
an exclusion at more than 99.9\% \CL\footnote{Regarding the exact
  value of this \CL\ see also the discussion of the (3+0) case in the
  next section.} We conclude that the data of LSND (using the result of
the analysis performed by the LSND collaboration) plays a very
significant role and that the global fit in a four-neutrino scenario
is much better than in the three-active neutrino case.

\section{Goodness of fit}
\label{sec:gof}

In the previous sections we have restricted ourselves to the {\it
  relative} comparison of the fit in the different mass schemes.
Here we discuss the absolute goodness of fit (GOF). A common way of
evaluating the GOF is to consider the absolute value of the
$\chi^2$-function at the best fit point. We are aware of the fact that
GOF-values obtained in this way are not very restrictive in a global
analysis with many data points like in our case. Therefore, we will
also consider in this section the quality of the fit in the
four-neutrino schemes (3+1), (2+2) and for the three active neutrino
case (3+0) for each of the different data sub-sets separately.

As explained in Sec.~\ref{sec:SBLdata} we are not able to use any
information on the absolute value of $\chi^2_\Lsnd$ from the LSND
data.  However, let us note that the fit for LSND is expected to be
rather good.  In Ref.~\onlinecite{LSND2001} the $\chi^2$ for the fit
of the $L/E$ distribution to the decay-at-rest events of the LSND data
is given for two typical values of $\dml$ as $\chi^2 = 4.9$ and 5.8
for 8 \dof, which corresponds to a very good GOF of 77\% and 67\%,
respectively.  From Fig.~\ref{fig:global} one can see that for (3+1)
as well as for (2+2) the best fit point lies well inside the 99\% \CL\
region of LSND.  Therefore, we expect that the contribution of LSND
will not worsen the global fit significantly (see also
Tab.~\ref{tab:pulloff}).

In the following we evaluate the $\chi^2$-functions for all
experiments except LSND
\begin{equation}
    \chi^2_{\mathrm{global}-\Lsnd} \equiv \chi^2_\Sol +
    \chi^2_\Atm + \chi^2_\Nev
\end{equation}
at the global best fit point in (3+1) and the best fit point in (2+2):
\begin{equation}\label{eq:gofchi2}
    \begin{aligned}
	\text{(3+1)} &:\: & \chi^2_{\mathrm{global}-\Lsnd}
	&= 150.0/176~\text{\dof}, \\
	\text{(2+2)} &:\: & \chi^2_{\mathrm{global}-\Lsnd}
	&= 156.1/176~\text{\dof}.
    \end{aligned}
\end{equation}
The number of \dof\ is given by (see Eqs.~\eqref{eq:NNEV},
\eqref{eq:Nsol}, \eqref{eq:Natm}) $N_\Sol + N_\Atm + N_\Nev = 185$
minus 9 fitted parameters. Usually a fit is considered to be good if
the value of the $\chi^2$ is approximately equal to the number of
\dof.

The GOF implied by the $\chi^2$-values and the corresponding number of
\dof\ given in Eq.~\eqref{eq:gofchi2} would be excellent for both
schemes. However, one has to be careful in the interpretation of these
numbers. This $\chi^2$-test for the GOF is not a very restrictive
criterion in a global fit of different experiments with a large number
of data points and many parameters.  One reason is that in such a case
a given parameter is constrained often only by a small sub-set of the
data. The rest of the data (which can contain many data points) is
fitted perfectly by this parameter (because it is insensitive to it).
A discussion of this problem can be found in Ref.~\onlinecite{collins}
or in the context of solar neutrino analysis in
Refs.~\onlinecite{solGof}.

In order to obtain more insight in the quality of the global fit we
will consider the following quantities:
\begin{equation}
    \Delta \chi^2_\sigma = \chi^2_\sigma(\alpha) -
    (\chi^2_\sigma)_\mathrm{min} \,.
\end{equation}
Here $\chi^2_\sigma$ is the $\chi^2$-function of the data set $\sigma = \Sol$,
$\Atm$, $\Lsnd$, $\Nev$ and $(\chi^2_\sigma)_\mathrm{min}$ is the minimum of
$\chi^2_\sigma$. This quantity can be used to test if a given point $\alpha$
in the parameter space is in agreement with the data set $\sigma$. For
$\alpha$ we will use the best fit points from the global analysis for (3+1)
and (2+2), the local minimum in (2+2) and the point corresponding to the (3+0)
case. This approach is similar to the method proposed in
Ref.~\onlinecite{collins}.

\begin{table}[t] \centering
    \catcode`?=\active \def?{\hphantom{0}}
    \begin{tabular}{l@{\hspace{2mm}}|@{\hspace{2mm}}c@{\hspace{2mm}}|
	  @{\hspace{2mm}}l@{\hspace{2mm}}|@{\hspace{2mm}}
	  c@{\hspace{2mm}}c@{\hspace{2mm}}c@{\hspace{2mm}}c} \hline\hline
	data set & \dof & parameters & (3+1)
	& (2+2)$_\mathrm{best}$ & (2+2)$_\mathrm{local}$ & (3+0) \\
	\hline
	solar & 3 & $\dms,\,\theta_\Sol,\,\eta_s$ 
	& $0.0$ & $10.7$ & $?1.6$ & $?0.0$ \\
	atmospheric & 4 & $\dma,\,\theta_\Atm,\,\eta_s,\,d_\mu$
	& $0.0$ & $?0.2$ & $11.5$ & $?0.3$ \\
	LSND & 2 & $\dml,\, \theta_\Lsnd$ 
	& $3.0$ & $?0.7$ & $?0.7$ & $29.0$ \\
	NEV & 2/3 & $\theta_\Lsnd,\,d_e,\,d_\mu$ 
	& $8.8$ & $?3.7$ & $?4.1$ & $?2.3$ \\
	\hline\hline
    \end{tabular} 
    \caption{\label{tab:pulloff}%
      $\Delta\chi^2$ for the different data sets of the best fit points in
      (3+1) and (2+2), the local minimum in (2+2) and for the (3+0) case (see
      text for details). Also shown is the number of \dof\ and the
      corresponding parameters for each data set.}
\end{table}

Let us discuss the results of this analysis, which are shown in
Tab.~\ref{tab:pulloff}. One can see that solar and atmospheric data
are in perfect agreement with the global best fit point in (3+1).  The
reason is that in this case both effects are explained by active
neutrino oscillations, which is preferred by the data. Also a $\Delta
\chi^2 = 3$ for LSND is in good agreement; the best fit point lies
within the 90\% \CL\ region for the two parameters $\dml$ and $\sql$.
However, the (3+1) best fit point gives a rather bad fit to the SBL
experiments finding no evidence of oscillations: a value of $\Delta
\chi^2 = 8.8$ for 2 \dof\ ($d_e$ and $d_\mu$) is ruled out at 98.7\%
\CL.

Regarding the (2+2) scheme we observe some problems in the fit of
solar or atmospheric data: At the best fit (2+2) solution we obtain
for solar data $\Delta \chi^2 = 10.7$ for 3 \dof, which is ruled out
at 98.7\% \CL, while the fit of the other data sets ATM, LSND and NEV
is very good. The reason for the problems in the solar data is that
the best fit for (2+2) prefers a large value of $\eta_s$
(corresponding to the LOW solution). This implies a large component of
the sterile neutrino in solar oscillations, which gives a bad fit. In
the local minimum for (2+2) --~which is marked with a triangle in the
upper right panel of Fig.~\ref{fig:global} and corresponds to the LMA
solution~-- the fit of solar data is very good, whereas atmospheric
data gives a $\Delta \chi^2 = 11.5$ for 4 \dof, which is ruled out at
97.8\% \CL. In this case the bad fit is a consequence of the large
sterile component in atmospheric oscillations implied by the small
value of $\eta_s$.  The interesting shape of $\chi^2_\mathrm{global}
(\eta_s)$ in (2+2), which disfavors equal sterile admixture in solar
and atmospheric oscillations, implies that {\it either} the solar {\it
  or} the atmospheric fit is bad in the (2+2) case, but never both.

From the last column in Tab.~\ref{tab:pulloff} one can see that all
experiments except LSND are in perfect agreement with the (3+0)
scenario. However, as expected the fit of LSND is very bad in this
case and yields a $\Delta \chi^2 = 29$ for 2 \dof. In the Gaussian
approximation implied by Eq.~\eqref{eq:chi2LSND} this would be ruled
out at an extremely high \CL.  Let us note that far from the allowed
region of LSND this approximation may not be completely justified.
However, it is evident that LSND data is in strong disagreement with
no oscillations. According to Table X of Ref.~\onlinecite{LSND2001}
the probability that the observed number of excess events is due to a
fluctuation of the expected background is between $7.8\times 10^{-6}$
and $1.8\times 10^{-3}$, depending on different selection criteria
applied to the data.

Some remarks are in order regarding this analysis for the NEV
experiments. These experiments do not see any evidence for
oscillations and hence, they obtain no information on a mass-squared
difference; only an upper bound on the oscillation amplitude can be
derived. Therefore, we consider $\Delta\chi^2_\Nev$ at a {\it fixed}
value\footnote{Note that the original analyses of the
  Bugey~\cite{bugey} and CDHS~\cite{CDHS} collaborations were performed
  in this way.} of $\dml$.  Hence, the $\Delta\chi^2$-values shown
in the table have to be evaluated for 3 \dof\ ($\thl,d_e,d_\mu$) in
the (2+2) scheme and for 2 \dof\ for (3+1) because of
Eq.~\eqref{eq:3+1rel}. Depending on the mass scheme we fix $\dml$ at
the best fit values given in Tab.~\ref{tab:globalbest}; for (3+0) we
use the best fit value of $\dml$ in the (3+1) scheme.  Although the
NEV experiments are in agreement with no oscillations, the value
$\Delta\chi^2_\Nev=2.3$ for (3+0) shows that a small contribution of
$\dml$ can improve the fit slightly.

Table~\ref{tab:pulloff} confirms the results of
Secs.~\ref{sec:sol+atm} and \ref{sec:atm+SBL}. A combination of only
solar and atmospheric data prefers the (3+1) schemes. Therefore, these
data are in perfect agreement with the global fit in (3+1), but fit
worse in (2+2). On the other hand, atmospheric and SBL data prefer the
(2+2) scheme; the (3+1) best fit point is somewhat in disagreement
with NEV data. This conflict between different data sets does not show
up in the $\chi^2$-values given in Eq.~\eqref{eq:gofchi2}, since the
coupling between the data sets is rather weak. As discussed in
Sec.~\ref{sec:paramSummary} only the parameter $\eta_s$ is common to
solar and atmospheric oscillations, only the parameter $d_\mu$ links
atmospheric and SBL data, while there is no direct coupling of solar
and SBL data. All the other 7 parameters can be adjusted to give a
good fit of the corresponding data set. The remaining conflict between
the data sets is completely washed out by the large number of data
points, which are fitted perfectly.

\section{Conclusions}
\label{sec:conclusions}

We have presented a unified global analysis of current neutrino
oscillation data within the framework of four-neutrino mass schemes,
paying attention to the inequivalent classes of (3+1) and (2+2)
models. We have included all data from solar and atmospheric neutrino
experiments, as well as information from short-baseline experiments
including LSND and the null-result oscillation experiments.
We have mapped the leptonic mixing matrix into a set of parameters in
such a way that they have a well-defined physical meaning in each data
set, independently of the mass scheme ((3+1) or (2+2)) considered. For
example, one of these parameters is $\eta_s$, the fraction of sterile
neutrinos in solar oscillations. Similarly $\theta_{24}$ describes the
fraction of $\nu_\mu$ in atmospheric oscillations and $\sql$ is
characterizing the SBL $\pnu{\mu}\to\pnu{e}$ amplitude. The fact that
it is possible to describe the results of any of the given set of
experiments in terms of physical quantities independent of the mass
scheme implies that none of the considered data sets (solar,
atmospheric, SBL appearance or SBL disappearance) can be used on its
own to discriminate between different mass spectra. This follows from
the approximation $\eta_e \approx 1$, which is motivated by the bounds
from reactor neutrino experiments, and from the strong hierarchy among
the mass-squared differences indicated by the data. We have shown how
the differences between the mass schemes manifest themselves only when
two or more data sets are combined.

We have found that combining only solar and atmospheric neutrino data,
the (3+1) type schemes are preferred, whereas atmospheric data in
combination with short-baseline data prefers (2+2) models.
By combining all data in a global analysis the (3+1) mass scheme gives
a slightly better fit than the (2+2) case, though all four-neutrino
schemes are presently acceptable. The LSND result disfavors the
three-active neutrino scenario with only $\dms$ and $\dma$ at 99.9\%
\CL\ with respect to the four-neutrino best fit model.
We have also performed a detailed analysis of the goodness of fit in
order to identify which sub-set of the data is in disagreement with
the best fit solution in a given mass scheme.

We have found that, in isolation, the LSND data play a crucial role in
suggesting the need for a four-neutrino scenario, at odds with every
other piece of information. The upcoming MiniBooNE
experiment~\cite{MiniBooNE} will test the very important result of
LSND in the near future. However, we have shown in this work that the
existing data cannot decide between (2+2) and (3+1) mass schemes in a
statistically significant way. Most likely this problem will remain
also if MiniBooNE would confirm the LSND result, and to resolve the
ambiguity more experimental information will be needed.

Such information could be provided by experiments with a high sensitivity to
the sterile component in solar and/or atmospheric neutrino oscillations. One
possibility to improve this sensitivity for atmospheric neutrinos could be the
consideration of neutral current events in atmospheric neutrino
experiments~\cite{Habig:2001ej}. Concerning solar neutrinos, we note that the
different oscillation solutions show very distinct behavior with respect to a
sterile component (see Fig.~\ref{fig:etas_sol}); hence, the identification of
the true solution is important. Moreover, improved measurements of neutral
current event rates may increase the sensitivity to sterile oscillations.
However, as shown in Ref.~\onlinecite{Barger:2002zs}, the information
obtainable from the neutral current measurements currently performed at SNO
will be limited because of the relatively large uncertainty in the flux of
solar $^8$B neutrinos.
On the other hand, more data on $\pnu{e}$ and/or $\pnu{\mu}$ SBL disappearance
probabilities could help to solve the (3+1) versus (2+2) puzzle. Especially
the existing bounds on $\pnu{\mu}$ disappearance are rather weak. This will be
improved by the MiniBooNE experiment, which -- beside testing the
$\nu_\mu\to\nu_e$ appearance channel -- will also provide a new measurement
for the $\nu_\mu$ survival probability~\cite{MiniBooNE}.

In view of the ambiguities implied by present data we are looking forward to
the results of future neutrino oscillation experiments, which may unravel the
secret behind the structure of the leptonic weak interaction.

\begin{acknowledgments}
    We thank P.~Huber and C.~Pe{\~n}a-Garay for many useful
    discussions and W.~Grimus for comments on a preliminary version of
    this paper.  We are very grateful to W.C.~Louis and G.~Mills for
    providing us with the table of the LSND likelihood function.  This
    work was supported by Spanish DGICYT under grant PB98-0693, by the
    European Commission RTN network HPRN-CT-2000-00148 and by the
    European Science Foundation network grant N.~86. T.~S. was
    supported by a fellowship of the European Commission Research
    Training Site contract HPMT-2000-00124 of the host group. M.~M.
    was supported by contract HPMF-CT-2000-01008.
\end{acknowledgments}


\begin{thebibliography}{99}

%%%%%%%%%%%%%%% solar %%%%%%%%%%%%%%%%%%%%

\bibitem{sksol} Super-Kamiokande Coll., Y. Fukuda {\it et al.},
  Phys. Rev. Lett. {\bf 81}, 1158 (1998);
  Erratum {\it ibid.} {\bf 81}, 4279 (1998)
  and {\it ibid}. {\bf 82}, 1810 and 2430 (1999);
  Y. Suzuki, Nucl. Phys. B (Proc. Suppl.) {\bf 77}, 35 (1999);
  S.~Fukuda {\it et al.}, hep-ex/0103032.

\bibitem{chlorine} B.T.~Cleveland {\it et al.},
  Astrophys. J. {\bf 496}, 505 (1998);
  R. Davis, Prog. Part. Nucl. Phys. {\bf 32}, 13 (1994);
  K. Lande, Talk given at \textit{Neutrino 2000},
  15--21 June 2000, Sudbury, Canada
  [\verb"http://nu2000.sno.laurentian.ca"].
  %
\bibitem{sage} SAGE Coll., D.N.~Abdurashitov {\it et al.},
  Phys. Rev. {\bf C60}, 055801 (1999);
  V. Gavrin, Talk given at \textit{Neutrino 2000},
  15--21 June 2000, Sudbury, Canada
  [\verb"http://nu2000.sno.laurentian.ca"].
  %
\bibitem{gallex_gno} GALLEX Coll., W.~Hampel {\it et al.},
  Phys. Lett. {\bf B447}, 127 (1999);
  E. Belloti, GNO Coll., Talk given at \textit{Neutrino 2000},
  15--21 June 2000, Sudbury, Canada
  [\verb"http://nu2000.sno.laurentian.ca"].
  %  \textit{(http://{\-}nu2000.{\-}sno.{\-}laurentian.{\-}ca)}.
  %
\bibitem{sno}
  Q.~R.~Ahmad {\it et al.}, SNO Coll., nucl-ex/0106015.

%%%%%%%%%%%%%%%% atmospheric %%%%%%%%%%%%%%%%%%%%%

\bibitem{atm-exp}
  Y. Fukuda \textit{et al.},
  Kamiokande Coll., Phys. Lett. B \textbf{335} (1994) 237;
  %
  R. Becker-Szendy \textit{et al.},
  IMB Coll., Nucl. Phys. B (Proc. Suppl.) \textbf{38} (1995) 331;
  %
  W.W.M. Allison \textit{et al.}, Soudan Coll.,
  Phys. Lett. B \textbf{449} (1999) 137.

\bibitem{skatm} Super-Kamiokande Coll.,
  Y. Fukuda \textit{et al.}, Phys. Rev. Lett. \textbf{81} (1998) 1562;
  C. McGrew in {\it Neutrino Telescopes 2001}, Venice, Italy, March 2001,
  to appear; T. Toshito in {\it Moriond 2001}, Les Arcs, France, March 2001,
  to appear.

\bibitem{macroOld} MACRO Coll.,
  M. Ambrosio \textit{et al.}, Phys. Lett. B \textbf{434} (1998) 451;
  M. Spurio {\it et al.}, hep-ex/0101019;
  B. Barish, Talk given at
  \textit{Neutrino 2000}, 15--21 June 2000, Sudbury, Canada
  [\verb"http://nu2000.sno.laurentian.ca"].

\bibitem{macroNew} A.~Surdo, Talk given at \textit{TAUP 2001},
  8--12 September 2001, Gran Sasso, Italy
  [\verb"http://www.lngs.infn.it/"].

%%%%%%%%%%%%%%%%%%%%%

\bibitem{LSND}
  C. Athanassopoulos \textit{et al.},
  LSND Coll., Phys. Rev. Lett. \textbf{77} (1996) 3082;
  \emph{ibid} \textbf{81} (1998) 1774;
  G. Mills, LSND Coll., Talk given at \textit{Neutrino 2000},
  15--21 June 2000, Sudbury, Canada
  [\verb"http://nu2000.sno.laurentian.ca"].

\bibitem{LSND2001}
  A.~Aguilar {\it et al.}, LSND Coll., hep-ex/0104049.

\bibitem{CPT}
  H.~Murayama and T.~Yanagida, hep-ph/0010178;
  G.~Barenboim \textit{et al.}, hep-ph/0108199.

\bibitem{ptv92}
  J.~T.~Peltoniemi, D.~Tommasini and J.~W.~F.~Valle,
  %``Reconciling dark matter and solar neutrinos,''
  Phys.\ Lett.\ B {\bf 298} (1993) 383.
  %%CITATION = PHLTA,B298,383;%%

\bibitem{pv93}
  J.~T.~Peltoniemi and J.~W.~F.~Valle,
  %``Reconciling dark matter, solar and atmospheric neutrinos,''
  Nucl.\ Phys.\ B {\bf 406}, 409 (1993)
  [hep-ph/9302316].
  %%CITATION = HEP-PH 9302316;%%

\bibitem{cm93}
  D.O. Caldwell and R.N. Mohapatra, Phys. Rev. D \textbf{48} (1993) 3259.

\bibitem{4nuModels}
  E. Ma and P. Roy, Phys. Rev. D \textbf{52} (1995) R4780;
  E.J. Chun \textit{et al.}, Phys. Lett. B \textbf{357} (1995) 608;
  J.J. Gomez-Cadenas and M.C. Gonzalez-Garcia,
  Z. Phys. C \textbf{71} (1996) 443;
  E. Ma, Mod. Phys. Lett. A \textbf{11} (1996) 1893;
  S. Goswami, Phys. Rev. D \textbf{55} (1997) 2931;
  %
  Q.Y. Liu and A. Smirnov, Nucl. Phys. B {\bf 524} 505;
  V. Barger, K. Whisnant and T. Weiler, Phys. Lett. B {\bf 427} (1998) 97;
  S. Gibbons, R.N. Mohapatra, S. Nandi and A. Raichoudhuri,
  Phys. Lett. B {\bf 430} (1998) 296; Nucl. Phys. B {\bf 524} (1998) 505;
  E.J. Chun, A. Joshipura and A. Smirnov, in {\sl Elementary Particle
    Physics: Present and Future} (World Scientific, 1996), ISBN 981-02-2554-7;
  P. Langacker, Phys. Rev. D {\bf 58} (1998) 093017;
  M. Bando and K. Yoshioka, Prog. Theor. Phys. {\bf 100} (1998) 1239;
  W. Grimus, R. Pfeiffer and T. Schwetz, Eur. Phys. J. C {\bf 13} (2000) 125;
  F. Borzumati, K. Hamaguchi and T. Yanagida, Phys. Lett. B {\bf 497} (2001) 259;
  K.~R.~Balaji, A.~Perez-Lorenzana and A.~Y.~Smirnov,
  Phys.\ Lett.\ B {\bf 509} (2001) 111;
  K.S.~Babu and R.N.~Mohapatra, hep-ph/0110243;
  S.~Goswami and A.S.~Joshipura, hep-ph/0110272.

\bibitem{4nuextra} R.~N.~Mohapatra, A.~Perez-Lorenzana and C.~A.~de S
  Pires,
  %``Neutrino mass, bulk Majoron and neutrinoless double beta decay,''
  Phys.\ Lett.\ B {\bf 491} (2000) 14.

\bibitem{Ioannisian:2001mu}
  A.~Ioannisian and J.~W.~F.~Valle,
  %``Light sterile neutrino from extra dimensions and four neutrino solutions to neutrino anomalies,''
  Phys.\ Rev.\ D {\bf 63} (2001) 073002;
  %%CITATION = PHRVA,D63,073002;%%
  [hep-ph/9911349]. %%CITATION = HEP-PH 9911349;%%

\bibitem{Hirsch:2000xe}
  M.~Hirsch and J.~W.~F.~Valle,
  %``Reconciling neutrino anomalies in a simple four-neutrino scheme with  R-parity violation,''
  Phys.\ Lett.\ B {\bf 495} (2000) 121
  [hep-ph/0009066].
  %%CITATION = HEP-PH 0009066;%%

\bibitem{giuntiwebp} Web-page of C.~Giunti,
  \verb"http://www.to.infn.it/~giunti/neutrino/" .

\bibitem{barger00}
  V. Barger \textit{et al.}, Phys. Lett. B \textbf{489} (2000) 345.
  %hep-ph/0008019


\bibitem{peres}
  O.L.G. Peres and A.Yu. Smirnov, Nucl. Phys. B {\bf 599} (2001) 3.

\bibitem{OY}
  N. Okada and O. Yasuda,
  Int. J. Mod. Phys. A \textbf{12} (1997) 3669.

\bibitem{barger98}
  V. Barger, S. Pakvasa, T.J. Weiler and K. Whisnant,
  Phys. Rev. D \textbf{58} (1998) 093016.

\bibitem{BGGS}
  S.M. Bilenky, C. Giunti, W. Grimus and T. Schwetz,
  Phys. Rev. D \textbf{60} (1999) 073007.

\bibitem{BGG}
  S.M. Bilenky, C. Giunti and W. Grimus,
  in \textit{Proceedings of Neutrino '96}, Helsinki, Finland, 13--19
  June 1996, p.\ 174,
  edited by K. Enqvist, K. Huitu and J. Maalampi (World Scientific,
  Singapore 1997);
  S.M. Bilenky, C. Giunti and W. Grimus, Eur. Phys. J. C \textbf{1} (1998) 247.

\bibitem{carlo}
  C. Giunti and M. Laveder, JHEP {\bf 0102} (2001) 001.

%<TeX_Marker>
\bibitem{GS}
  W. Grimus and T. Schwetz, Eur. Phys. J. C {\bf 20} (2001) 1 [hep-ph/0102252].

\bibitem{Maltoni:2001mt}
  M.~Maltoni, T.~Schwetz and J.~W.~F.~Valle,
  %``Cornering (3+1) sterile neutrino schemes,''
  Phys.\ Lett.\ B {\bf 518}, 252 (2001)
  [hep-ph/0107150].

\bibitem{sksterile}
  S.~Fukuda {\it et al.}, Super-Kamiokande Coll.,
  hep-ex/0103033;
  %
  Phys.\ Rev.\ Lett.\  {\bf 85}, 3999 (2000).

\bibitem{bahcall}
  J.~N.~Bahcall, M.~C.~Gonzalez-Garcia and C.~Pena-Garay,
  hep-ph/0106258.

\bibitem{Concha:2001uy}
  M.C.~Gonzalez-Garcia, M.~Maltoni and C.~Pe{\~n}a-Garay,
  %``Solar and atmospheric four-neutrino oscillations,''
  Phys.\ Rev.\ D {\bf 64} (2001) 093001
  [hep-ph/0105269].

\bibitem{Concha:2001zi}
  M.~C.~Gonzalez-Garcia, M.~Maltoni and C.~Pe{\~n}a-Garay,
  %``Update on solar and atmospheric four-neutrino oscillations,''
  hep-ph/0108073.

\bibitem{KARMEN}
  K. Eitel and B. Zeitnitz, KARMEN Coll., in
  \textit{Proceedings of Neutrino '98}, Takayama, Japan, 4-9 June 1998,
  Nucl. Phys. Proc. Suppl. \textbf{77}, 212 (1999); %hep-ex/9809007
  K. Eitel, KARMEN Coll., Talk given at
  \textit{Neutrino 2000}, 15--21 June 2000, Sudbury, Canada,
  hep-ex/0008002.

\bibitem{nomad}
  M. Mezzetto, NOMAD Coll., Nucl. Phys. B (Proc. Suppl.) {\bf 70} (1999) 214;
  L. Camilleri, NOMAD Coll., {\it A Status Report} (1999)
  [\verb"http://nomadinfo.cern.ch/Public/PUBLICATIONS/"].

\bibitem{bugey}
  B. Achkar \textit{et al.}, Bugey Coll., Nucl. Phys. B \textbf{434} (1995) 503.

\bibitem{CHOOZ}
  M. Apollonio \textit{et al.}, CHOOZ Coll.,
  Phys. Lett. B \textbf{466} (1999) 415.

\bibitem{CDHS}
  F. Dydak \textit{et al.}, CDHS Coll., Phys. Lett. B \textbf{134} (1984) 281.

\bibitem{Schechter:1980gr}
  J.~Schechter and J.~W.~F.~Valle,
  %``Neutrino Masses In SU(2) X U(1) Theories,''
  Phys.\ Rev.\ D {\bf 22} (1980) 2227.
  %%CITATION = PHRVA,D22,2227;%%

\bibitem{Schechter:1981gk}
  J.~Schechter and J.~W.~F.~Valle,
  %``Neutrino Oscillation Thought Experiment,''
  Phys.\ Rev.\ D {\bf 23} (1981) 1666.
  %%CITATION = PHRVA,D23,1666;%%


\bibitem{CP}
  S.M.~Bilenky, C.~Giunti and W.~Grimus, Phys. Rev. D {\bf 58} (1998) 033001;
  V.~Barger {\it et al.}, Phys. Rev. D {\bf 59} (1999) 113010;
  K.~Dick {\it et al.}, Nucl. Phys. B {\bf 562} (1999) 29;
  A.~Donini {\it et al.}, Nucl. Phys. B {\bf 574} (2000) 23;
  A.~Kalliomaki, J.~Maalampi and M.~Tanimoto,
  Phys. Lett. B {\bf 469} (1999) 179;
  T.~Hattori, T.~Hasuike and S.~Wakaizumi, Phys. Rev. D {\bf 62} (2000) 033006;
  A.~Donini and D.~Meloni, hep-ph/0105089.

\bibitem{farzan}
  Y.~Farzan, O.L.G.~Peres and A.Yu.~Smirnov, 
  Nucl.\ Phys.\ B {\bf 621} (2001) 59 [hep-ph/0105105].

\bibitem{nu0bdec}
  S.M. Bilenky, S. Pascoli and S.T. Petcov, hep-ph/0104218;
  A.~Kalliomaki and J.~Maalampi,
  Phys.\ Lett.\ B {\bf 484} (2000) 64 [hep-ph/0003281];
  S.~M.~Bilenky, C.~Giunti, W.~Grimus, B.~Kayser and S.~T.~Petcov,
  Phys.\ Lett.\ B {\bf 465} (1999) 193 [hep-ph/9907234].

\bibitem{PaloV}
  F.~Boehm {\it et al.}, Palo Verde Coll.,
  %``Final results from the Palo Verde neutrino oscillation experiment,''
  Phys.\ Rev.\ D {\bf 64} (2001) 112001
  [hep-ex/0107009].
  %%CITATION = HEP-EX 0107009;%%


\bibitem{Gonzalez-Garcia:2000aj}
  M.~C.~Gonzalez-Garcia, P.~C.~de Holanda, C.~Pena-Garay and J.~W.~F.~Valle,
  %``Status of the MSW solutions of the solar neutrino problem,''
  Nucl.\ Phys.\ B {\bf 573} (2000) 3
  [hep-ph/9906469].
  %%CITATION = HEP-PH 9906469;%%

\bibitem{LSNDpriv}
  W.C.~Louis and G.~Mills, LSND Coll., private communication.

\bibitem{pdg}
  D.E.~Groom \textit{et al.}, Review of Particle Physics,
  Eur. Phys. J. C \textbf{15} (2000) 1.

\bibitem{Dooling:2000sg}
  D.~Dooling, C.~Giunti, K.~Kang and C.~W.~Kim,
  %``Matter effects in four-neutrino mixing,''
  Phys.\ Rev.\ D {\bf 61} (2000) 073011
  [hep-ph/9908513].

\bibitem{Giunti:2000wt}
  C.~Giunti, M.~C.~Gonzalez-Garcia and C.~Pe{\~n}a-Garay,
  %``Four-neutrino oscillation solutions of the solar neutrino problem,''
  Phys.\ Rev.\ D {\bf 62} (2000) 013005
  [hep-ph/0001101].

\bibitem{BGGSbbn}
  S.~M.~Bilenky, C.~Giunti, W.~Grimus and T.~Schwetz,
  %``Four-neutrino mixing and big-bang nucleosynthesis,''
  Astropart.\ Phys.\  {\bf 11} (1999) 413 [hep-ph/9804421].
  %%CITATION = HEP-PH 9804421;%%

\bibitem{GMPV}
  M.~C.~Gonzalez-Garcia, M.~Maltoni, C.~Pena-Garay and J.~W.~F.~Valle,
  %``Global three-neutrino oscillation analysis of neutrino data,''
  Phys.\ Rev.\ D {\bf 63} (2001) 033005 [hep-ph/0009350];
  %%CITATION = HEP-PH 0009350;%%
  %
  N. Fornengo, M.C. Gonzalez-Garcia and J.~W.~F. Valle, Nucl. Phys. B {\bf
    580} (2000) 58;
  %
  M.~C.~Gonzalez-Garcia, H.~Nunokawa, O.~L.~G.~Peres and J.~W.~F.~Valle,
  %``Active-active and active-sterile neutrino oscillation solutions to the  atmospheric neutrino anomaly,''
  Nucl.\ Phys.\ B {\bf 543} (1999) 3 [hep-ph/9807305].
  %%CITATION = HEP-PH 9807305;%%

\bibitem{atm4nuFit}
  G.~L.~Fogli, E.~Lisi and A.~Marrone,
  %``Four-neutrino oscillation solutions of the atmospheric neutrino  anomaly,''
  Phys.\ Rev.\ D {\bf 63} (2001) 053008 [hep-ph/0009299]; O.~Yasuda,
  hep-ph/0006319.

\bibitem{eitel}
  K.~Eitel, New Jour.\ Phys.\  {\bf 2} (2000) 1
  [hep-ex/9909036].

\bibitem{BNL}
  L. Borodovsky {\it et al.}, Phys. Rev. Lett. \textbf{68} (1992) 274.

\bibitem{CCFR}
  A. Romosan {\it et al.}, Phys. Rev. Lett. \textbf{78} (1997) 2912;
  I.E.~Stockdale {\it et al.}, Phys. Rev. Lett. \textbf{52} (1984) 1384;
  Z. Phys. C \textbf{53} (1985) 53.

\bibitem{bahcallNew}
  J.~N.~Bahcall, M.~C.~Gonzalez-Garcia and C.~Pena-Garay,
  hep-ph/0111150.

\bibitem{collins}
  J.~C.~Collins and J.~Pumplin, hep-ph/0105207.

\bibitem{solGof}
  P.~Creminelli, G.~Signorelli and A.~Strumia,
  JHEP {\bf 0105} (2001) 052 [hep-ph/0102234];
  %
  M.~V.~Garzelli and C.~Giunti, hep-ph/0108191;
  %
  P.I.~Krastev and A.Yu.~Smirnov, hep-ph/0108177.

\bibitem{MiniBooNE}
  A. Bazarko, MiniBooNE Coll., Talk given at
  \textit{Neutrino 2000}, 15--21 June 2000, Sudbury, Canada
  [\verb"http://nu2000.sno.laurentian.ca"].

\bibitem{Habig:2001ej} 
A.~Habig, Super-Kamiokande Coll.,
%``Discriminating between nu/mu <--> nu/tau and nu/mu <--> nu/sterile in  
%atmospheric nu/mu oscillations with the Super-Kamiokande detector,''
hep-ex/0106025.
%%CITATION = HEP-EX 0106025;%%

\bibitem{Barger:2002zs}
V.~D.~Barger, D.~Marfatia and K.~Whisnant,
%``Unknowns after the SNO charged-current measurement,''
Phys.\ Rev.\ Lett.\  {\bf 88} (2002) 011302 [hep-ph/0106207].
%%CITATION = HEP-PH 0106207;%%

\end{thebibliography}
\end{document}